\documentclass[prd,amsmath,notitlepage,superscriptaddress,twocolumn,nofootinbib]{revtex4-1}
\usepackage{graphicx}
\usepackage{float}
\usepackage{tensor}
\usepackage{epstopdf,cancel}
\usepackage{epsf,latexsym,bbm,euscript}
\usepackage{amssymb,amsmath}
\usepackage{mathtools} 
\usepackage{times,graphics}
\usepackage{soul,xcolor}
\usepackage{mathtools}

\usepackage{enumitem}

\def\6{{\langle}}
\def\9{{\rangle}}
\newcommand{\defeq}{\vcentcolon=}
\newcommand{\eqdef}{=\vcentcolon}

\newcommand{\be}{\begin{equation}}
\newcommand{\ee}{\end{equation}}
\newcommand{\ba}{\begin{eqnarray}}
\newcommand{\ea}{\end{eqnarray}}

\def\half{{\tfrac{1}{2}}}

\def\pad{{\partial}}

\def\sg{\textsl{g}}

 \def\eE{\EuScript{E}}
 \def\eF{\EuScript{F}}
 \def\eL{\EuScript{L}}
\def\cO{\mathcal{O}}

\def\mE{\mathfrak{E}}

\newcommand{\sumj}{\sum\limits_{j \geqslant \frac{1}{2}}^\infty}
\newcommand{\overbar}[1]{\mkern 1.5mu\overline{\mkern-1.5mu#1\mkern-1.5mu}\mkern 1.5mu}

\usepackage{csquotes}
\usepackage{cases}

\makeatletter
\renewenvironment{numcases}[1]%
{$$\numc@opts \setbox\z@\hbox
  {\advance\c@equation\@ne\def\@currentlabel{\p@equation\theequation}
  $\displaystyle {#1\null}\m@th$}%
 \numc@setsub
 \setbox\tw@\vbox\bgroup
  \stepcounter{equation}\def\@currentlabel{\p@equation\theequation}%
  \global\@eqnswtrue\m@th \everycr{}\tabskip\numc@left\let\\\@eqncr
  \halign to\dimen@ii \bgroup \kern\wd\z@ \kern14\p@ 
    \tabskip\z@skip \global\@eqcnt\@ne $\displaystyle{##}$\hfil
   &\global\@eqcnt\tw@ \skip@10\p@ \advance\skip@\tw@\arraycolsep \hskip\skip@
    ##\unskip\hfil\tabskip\@centering
   &\global\@eqcnt\thr@@\hbox to\z@\bgroup\hss##\egroup\tabskip\z@skip\cr
}{\@@eqncr \egroup 
 \global\advance\c@equation\m@ne
 \unskip\unpenalty\unskip\unpenalty \setbox\z@\lastbox 
\ifvbox\z@
\setbox\tw@\box\z@
\setbox\z@\lastbox
 \nointerlineskip \copy\z@ 
 \nointerlineskip \copy\tw@ 
\else
 \nointerlineskip \copy\z@ 
\fi
 \global\dimen@i\wd\z@
 \setbox\z@\hbox{\hskip-\numc@left\unhbox\z@}
 \ifdim \wd\z@<\dimen@i \global\dimen@i\wd\z@ \fi
\egroup
\hbox to\dimen@ii{\m@th 
  \hskip\numc@left
  \hbox to\dimen@i{$\displaystyle \box\z@ 
    \dimen@\ht\tw@ \advance\dimen@\dp\tw@ 
    \left\{\vcenter to\dimen@{\vfil}\right.\n@space 
    $\hfil}\hskip\@centering 
  \kern-\dimen@ii 
  $\vcenter{\box\tw@}$
 }
\numc@resetsub
$$\global\@ignoretrue}
\makeatother

\usepackage{url,hyperref}
\hypersetup{colorlinks,linkcolor={blue!55!black},citecolor={red!45!black},urlcolor={blue!45!black},breaklinks=true}

\begin{document}

\title{Spherically symmetric black holes in metric gravity}

\author{Sebastian Murk}
\email{sebastian.murk@mq.edu.au}
\affiliation{Department of Physics and Astronomy, Macquarie University, Sydney, New South Wales 2109, Australia}
\affiliation{Sydney Quantum Academy, Sydney, New South Wales 2006, Australia}

\author{Daniel R.\ Terno}
\email{daniel.terno@mq.edu.au}
\affiliation{Department of Physics and Astronomy, Macquarie University, Sydney, New South Wales 2109, Australia}

\begin{abstract}
	The existence of black holes is one of the key predictions of general relativity (GR) and therefore a basic consistency test for modified theories of gravity. In the case of spherical symmetry in GR the existence of an apparent horizon and its regularity is consistent with only two distinct classes of physical black holes. Here we derive constraints that any self-consistent modified theory of gravity must satisfy to be compatible with their existence. We analyze their properties and illustrate characteristic features using the Starobinsky model. Both of the GR solutions can be regarded as zeroth-order terms in perturbative solutions of this model. We also show how to construct nonperturbative solutions without a well-defined GR limit.
\end{abstract}

\maketitle

\section{Introduction} \label{sec:Introduction}

General Relativity (GR), one of the two pillars of modern physics,  is the simplest member of the family of metric theories of gravity. It is the only theory that is derived from an invariant that is linear in second derivatives of the metric. However, interpretations of astrophysical and cosmological data as well as theoretical considerations \cite{cdl:11,f-R} encourage us to consider GR as the low-energy limit of some effective theory of quantum gravity \cite{dh:15,burgess:04,hv:book}. Extended theories of gravity, such as metric theories that involve higher-order invariants of the Riemann tensor, metric-affine theories, and theories with torsion, include additional terms in the action functional. Here we focus on metric modified theories of gravity (MTG).

A prerequisite for the validity of any proposed generalization of GR is that it must be compatible with current astrophysical and cosmological data. In particular, a viable candidate theory must provide a model to describe the observed astrophysical black hole candidates. Popular contemporary models describe them as ultra-compact objects with or without a horizon \cite{bh-map}. While there is a considerable diversity of opinions on what exactly constitutes a black hole, the presence of a trapped region --- a domain of spacetime from which nothing can escape --- is its most commonly accepted characteristic \cite{curiel}. A trapped spacetime region that is externally bounded by an apparent horizon is referred to as physical black hole (PBH) \cite{f:14}. A PBH may contain other features of black hole solutions of classical GR, such as an event horizon or singularity, or it may be a singularity-free regular black hole. To be of physical relevance, the apparent horizon must form in finite time according to a distant observer \cite{bmmt:19}.

It is commonly accepted that curvature invariants, such as the Ricci and Kretschmann scalar, are finite at the apparent horizon. When expressed mathematically, the requirements of regularity and finite formation time provide the basis for a self-consistent analysis of black holes. In spherical symmetry (to which we restrict our considerations here), this allows for a comprehensive classification of the near-horizon geometries. There are only two classes of solutions labeled by $k=0$ and $k=1$, where the value of $k$ reflects the scaling behavior of particular functions of the components of the energy-momentum tensor (EMT) near the apparent horizon. The properties of the near-horizon geometry lead to the identification of a unique scenario for black hole formation \cite{bmmt:19,mt:20} that involves both types of PBH solutions. We summarize its main results in Sec.~\ref{sec:GR}.

Understanding the true nature of the observed ultra-compact objects requires detailed knowledge of the black hole models, their alternatives, as well as the observational signatures of both classes of solutions in GR and extended theories of gravity \cite{cp-lrr:19,bh-map}. Vacuum black hole solutions exist in a variety of MTG \cite{cdl:11,f-R,ckms:18}. On the other hand, these theories are also used to construct models of horizonless ultra-compact objects. A generic property among some of them is the absence of horizon formation in the final stage of the collapse \cite{hr:17}.

Even the simplest MTG require perturbative treatment due to the mathematical complexity inherent to the higher-order nature of the equations \cite{f-R,sst:08,mvc:20}. We briefly review the relevant formalism and its relationship to the self-consistent approach  in Sec.~\ref{sec:modGravEqs}. In Sec.~\ref{sec:MTG}, we derive a set of conditions necessary for the existence of a PBH in an arbitrary metric MTG. The solutions are presented as expansions in the coordinate distance from the apparent horizon and do not require a GR solution as the zeroth-order perturbative solution of a MTG. Using the Starobinsky model \cite{f-R,staro} (Sec.~\ref{sec:Sgen}) we demonstrate the application of the general results, illustrating the well-known features of matching solutions of systems of partial differential equations of different orders \cite{f-R,sst:08,mvc:20}: we find that the two classes of GR solutions can be regarded as zeroth-order perturbative solutions of this MTG, and identify a MTG solution without a well-defined GR limit.

\section{Modified gravity field equations in spherical symmetry} \label{sec:modGravEqs} \subsection{General considerations}

We work in the framework of semiclassical gravity, use classical notions (e.g.\ metric, horizons, trajectories), and describe dynamics via the modified Einstein equations.
We do not make any assumptions about the underlying reason for modifications of the bulk part of the gravitational Lagrangian density, but organize it according to powers of derivatives of the metric as commonly done in effective field theories \cite{dh:15,burgess:04,pad:11}, i.e.\
\begin{align}
	\eL_\mathrm{g}\sqrt{-\sg}&= \frac{ M_\mathrm{P}^2}{16\pi}\big(R+\lambda \eF(\tensor{\sg}{^\mu^\nu}, \tensor{R}{_\mu_\nu_\rho_\sigma})\big)\nonumber \\
	&=\frac{ M_\mathrm{P}^2}{16\pi}R + a_1 \tensor{R}{_\mu_\nu} \tensor{R}{^\mu^\nu} + a_2 R^2 + a_3 \tensor{R}{_\mu_\nu_\rho_\sigma} \tensor{R}{^\mu^\nu^\rho^\sigma} + \cdots , \label{eq:eftL}
\end{align}
where $M_\mathrm{P}$ is the Planck mass that we set to one in what follows, the cosmological constant was omitted, and the coefficients $a_1$, $a_2$, $a_3$ are dimensionless. The dimensionless parameter $\lambda$ is used to organize the perturbative analysis and set to one at the end of the calculations. Many popular models belong to the class of $\mathfrak{f}(R)$ theories, where $\eL_\mathrm{g}\sqrt{-\sg}=\mathfrak{f}(R)$. The prototypical example is the Starobinsky model with $\eF=\varsigma R^2$, $\varsigma=16 \pi a_2/M^2_\mathrm{P}$.

Varying the gravitational action results in
\begin{align}
	\tensor{G}{_\mu_\nu} + \lambda \tensor{\EuScript{E}}{_\mu_\nu} = 8 \pi \tensor{T}{_\mu_\nu} , \label{eq:mEFE}
\end{align}
where $\tensor{G}{_\mu_\nu}$ is the Einstein tensor, the terms $\tensor{\EuScript{E}}{_\mu_\nu}$ result from the variation of $\eF(\tensor{\sg}{^\mu^\nu}, \tensor{R}{_\mu_\nu_\rho_\sigma})$, and $\tensor{T}{_\mu_\nu} \equiv \6 \tensor{\hat{T}}{_\mu_\nu}\9_\omega$ denotes the expectation value of the renormalized EMT. We do not make any specific assumptions about the state $\omega$.

In fact, apart from imposing spherical symmetry, we assume only that
(i) an apparent horizon is formed in finite time of a distant observer;
(ii) it is regular, i.e.\ the scalars
$\mathrm{T}\defeq \tensor{T}{^\mu_\mu} = {R} / 8\pi + \cO(\lambda)$ and
$\mathfrak{T}\defeq \tensor{T}{^\mu^\nu}\tensor{T}{_\mu_\nu} = \tensor{R}{^\mu^\nu}\tensor{R}{_\mu_\nu} / 64 \pi^2 + \cO(\lambda^2)$
are finite at the horizon.

A general spherically symmetric metric in Schwarzschild coordinates is given by
\begin{align}
	ds^2 = - e^{2h(t,r)} f(t,r) dt^2 + f(t,r)^{-1} dr^2 + r^2 d\Omega , \label{eq:metric}
\end{align}
where $r$ denotes the areal radius. The Misner\textendash{}Sharp mass \citep{cwM.dhS.1964,aphor} $C(t,r)$ is invariantly defined via
\begin{align}
	1 - C(t,r)/r \defeq \partial_\mu r \partial^\mu r , \label{eq:MSmass}
\end{align}
and thus the function $f(t,r) = 1 - C(t,r) / r$ is invariant under general coordinate transformations. For a Schwarzschild black hole $C = 2M$. We use the definition of Eq.~\eqref{eq:MSmass} for consistency with the description of solutions in higher-dimensional versions of GR. The apparent horizon is located at the Schwarzschild radius $r_\sg(t)$ that is the largest root of $f(t,r) = 0$ \cite{aphor}.

The Misner\textendash{}Sharp mass of a PBH can be represented as
\begin{align}
	C = r_\sg(t) + W(t,r-r_\sg) ,
\end{align}
where the definition of the apparent horizon implies
\begin{align}
	W(t,0) = 0 \; \; , \; \; W(t,x) < x,
\end{align}
and $x\defeq r-r_g$ is the coordinate distance from the apparent horizon.

The modified Einstein equations take the form
\begin{align}
	& f r^{-2} e^{2h} \partial_r C + \lambda \tensor{\EuScript{E}}{_t_t} = 8 \pi \tensor{T}{_t_t} , \label{eq:mEFEtt} \\
	& r^{-2} \partial_t C + \lambda \tensor{\EuScript{E}}{_t^r} = 8 \pi \tensor{T}{_t^r} , \label{eq:mEFEtr} \\
	& 2 f^2 r^{-1} \partial_r h - f r^{-2} \partial_r C + \lambda \tensor{\EuScript{E}}{^r^r} = 8 \pi \tensor{T}{^r^r} . \label{eq:mEFErr}
\end{align}

The notation
\begin{align}
	\tensor{\tau}{_t} \defeq e^{-2h} \tensor{T}{_t_t} \; \; , \; \; \tensor{\tau}{_t^r} \defeq e^{-h} \tensor{T}{_t^r} \; \; , \; \; \tensor{\tau}{^r} \defeq \tensor{T}{^r^r}
	\label{eq:tauFunctions}
\end{align}
is useful in dealing with equations in both GR and MTG.

Regularity of the apparent horizon is expressed as a set of conditions on the potentially divergent parts of the scalars $\mathrm{T}$ and $\mathfrak{T}$. In spherical symmetry $T^\theta_{\,\theta}\equiv T^\phi_{\,\phi}$ and we assume that it is finite as in GR \cite{bmmt:19}. The constraints can therefore be represented mathematically as
\begin{align}
	\mathrm{T} &= \left( \tensor{\tau}{^r} - \tensor{\tau}{_t} \right) / f \; \to \; g_1(t) f^{k_1} , \label{regT} \\
	\mathfrak{T} &= \left( (\tensor{\tau}{_t})^2 - 2 (\tensor{\tau}{_t^r})^2 + (\tensor{\tau}{^r})^2 \right) / f^2 \; \to \; g_2(t) f^{k_2} , \label{regT2}
\end{align}
for some $g_{1,2}(t)$ and $k_{1,2} \geqslant 0$. There are \textit{a priori} infinitely many solutions that satisfy these constraints. After reviewing the special case of GR and presenting the two admissible solutions we discuss this behavior in Sec.~\ref{sec:MTG}.

Many useful results can be obtained by means of comparison of various quantities written in Schwarzschild coordinates $(t,r)$ with their counterpart expressions written using the ingoing $v$ or outgoing $u$ null coordinate and the same areal radius $r$. Using $(v,r)$ coordinates,
\begin{align}
	dt=e^{-h}(e^{h_+}dv- f^{-1}dr), \label{intf}
\end{align}
is particularly fruitful. EMT components in $(v,r)$ and $(t,r)$ coordinates are related via
\begin{align}
	& \tensor{\theta}{_v} \defeq e^{-2h_+} \tensor{\Theta}{_v_v} = \tensor{\tau}{_t} , \label{thev} \\
	& \tensor{\theta}{_v_r} \defeq e^{-h_+} \tensor{\Theta}{_v_r} = \left( \tensor{\tau}{_t^r} - \tensor{\tau}{_t} \right) / f , \label{thevr} \\
	& \tensor{\theta}{_r} \defeq \tensor{\Theta}{_r_r} = \left( \tensor{\tau}{^r} + \tensor{\tau}{_t} - 2 \tensor{\tau}{_t^r} \right) / f^2 , \label{ther}
\end{align}
where $\tensor{\Theta}{_\mu_\nu}$ labels EMT components in $(v,r)$ coordinates.

\subsection{Perturbative expansion}

From a formal perspective the pure GR case can be described as a system of field equations \cite{wald:84}
\begin{align}
	\mE(\bar{\sg},\bar{T}) = 0 ,
\end{align}
where the EMT $\bar{T}$ and metric $\bar{\sg}$ near the apparent horizon are described in a spherically symmetric setting in Sec.~\ref{sec:GR}. It is then usually assumed that any solution
\begin{align}
	\mE_\lambda(\sg_\lambda, T_\lambda) = 0
\end{align}
of the MTG belongs to a one-parameter family of analytic solutions \cite{sst:08,mvc:20}. The EMT $T_\lambda$ depends on $\lambda$ through the metric $\sg_\lambda$, and potentially also through effective corrections resulting from perturbative corrections to the modified field equations Eqs.~\eqref{eq:mEFEtt}--\eqref{eq:mEFErr}. The self-consistent approach is based on the assumption of at least continuity of the curvature invariants, but uses the Schwarzschild coordinate system where the metric is discontinuous \cite{bmmt:19,mt:20}. Imposing the requirement of regularity then allows to identify the valid black hole solutions, whose analytic properties become apparent once they are written in their \enquote{natural} coordinate system \cite{t:19}.

The field equations are supplemented by a set of initial and boundary conditions or constraints. Higher-order terms in the action lead to higher-order equations. Even $\mathfrak{f}(R)$ theories already result in systems with fourth-order metric derivatives. However, it is worth pointing out that the unperturbed solution may not satisfy the boundary conditions since its corresponding equations do not involve the higher-order derivatives \cite{mvc:20,bc-18}.

For our purposes it suffices to restrict all considerations to first-order perturbation theory. In any given theory higher-order contributions can be successfully evaluated. There are methods to produce a consistent hierarchy of the higher-order terms and deal with additional degrees of freedom that result from the presence of derivatives of order higher than two. Nevertheless, including terms of order $\mathcal{O}(\lambda^2)$ and higher may not be justified without detailed knowledge of the relative importance of all possible terms in the effective Lagrangian and the cut-off scale that is used to derive it.

Spherical symmetry prescribes the form of the metric for all values of $\lambda$. We assume that there is a solution of  Eq.~\eqref{eq:mEFE}  with the two metric functions $C_\lambda$ and $h_\lambda$. To avoid spurious divergences we use the physical value of $r_\sg(t)$ that corresponds to the perturbed metric $\sg_\lambda$, $C_\lambda(r_\sg,t)=r_\sg$. We set
\begin{align}
	& C_\lambda \eqdef r_\sg(t)+ \bar{W}(t,r) + \lambda \Sigma(t,r) , \label{eq:pcSigma}\\
	& h_\lambda \eqdef \bar{h}(t,r) + \lambda \Omega(t,r) , \label{eq:pcOmega}
\end{align}
and define $\bar C \defeq r_\sg+\bar W$. Similarly, the EMT  $T_\lambda \equiv T$ is decomposed as
\begin{align}
	\tensor{T}{_\mu_\nu} \eqdef \tensor{\bar{T}}{_\mu_\nu} + \lambda \tilde{T},
\end{align}
where $\bar{T}$ is extracted from $\mE \big( \bar{\sg} [r_\sg,\bar{W}, \bar{h}] , \bar{T} \big)=0$.

The perturbative terms must satisfy the boundary conditions
\begin{align}
	& ~ \Sigma(t,0) = 0 , \\
	& \lim_{r \to r_\sg} \Omega(t,r) / \bar{h}(t,r) = \mathcal{O}(1) , \label{hdiver}
\end{align}
where the first condition follows from the definition of the Schwarzschild radius, and the perturbation can be treated as small only if the divergence of $\Omega$ is not stronger than that of $\bar{h}$. Substituting $C_\lambda$ and $h_\lambda$ into Eq.~\eqref{eq:mEFE} and keeping only the first-order terms in $\lambda$ results in
\begin{align}
	\tensor{\bar{G}}{_\mu_\nu} + \lambda \tensor{\tilde{G}}{_\mu_\nu} + \lambda \tensor{\bar{\EuScript{E}}}{_\mu_\nu} = 8 \pi \left( \tensor{\bar{T}}{_\mu_\nu} + \lambda \tensor{\tilde{T}}{_\mu_\nu} \right) , \label{MGscheme}
\end{align}
where $\tensor{\bar{G}}{_\mu_\nu} \equiv \tensor{G}{_\mu_\nu}[r_\sg, \bar{W}, \bar{h}]$, $\tensor{\tilde{G}}{_\mu_\nu}$ is the first-order term in the Taylor expansion in $\lambda$ where each monomial involves either $\Sigma$ or $\Omega$, and $\tensor{\bar{\EuScript{E}}}{_\mu_\nu} \equiv \tensor{\EuScript{E}}{_\mu_\nu}[r_\sg,\bar{W}, \bar{h}]$, i.e.\ the modified gravity terms are functions of the unperturbed solutions.

The explicit form of the equations can be obtained as follows. First note that
\begin{align}
	e^{2h}= e^{2 \bar{h}} \left( 1 + 2 \lambda \Omega \right)  +\mathcal{O}(\lambda^2) . \label{hdiver2}
\end{align}
We introduce the splitting $\tau = \bar{\tau} + \lambda \tilde{\tau}$ such that, for instance, the EMT terms of the $tt$ equation can be written as
\begin{align}
	\tensor{\bar{T}}{_t_t} + \lambda \tensor{\tilde{T}}{_t_t} &= e^{2 \bar{h}} \left( 1 + 2 \lambda \Omega \right) \left( \tensor{\bar{\tau}}{_t} + \lambda \tensor{\tilde{\tau}}{_t} \right) \\
	&= e^{2 \bar{h}} \bigl( \tensor{\bar{\tau}}{_t} + \lambda \left( 2 \Omega \tensor{\bar{\tau}}{_t} + \tensor{\tilde{\tau}}{_t} \right) \bigr) + \mathcal{O}(\lambda^2) ,
\end{align}
with $\tensor{T}{_t^r}$ and $\tensor{T}{^r^r}$ expanded analogously. The regularity conditions Eqs.~\eqref{regT} and \eqref{regT2} imply that $\tilde{\tau}$ terms should either have the same behavior as their $\bar{\tau}$ counterparts when $r \to r_\sg$, or go to zero faster.

Consequently, the schematic of Eq.~\eqref{MGscheme} implies
\begin{align}
	&	\tensor{\bar{G}}{_t_t} = \frac{e^{2 \bar{h}}}{r^3} \left( r - \bar{C} \right) \partial_r \bar{C} , \\
	& \tensor{\tilde{G}}{_t_t} = \frac{e^{2 \bar{h}}}{r^3} \left[ - \Sigma \partial_r \bar{C} + \left( r - \bar{C} \right)
 \left( 2 \Omega \partial_r \bar{C} + \partial_r \Sigma \right) \right] ,
\end{align}
and thus the explicit form of Eq.~\eqref{eq:mEFEtt} is
\begin{align}
	- \Sigma \partial_r \bar{C} + \left( r - \bar{C} \right) \partial_r \Sigma + r^3 e^{- 2 \bar{h}} \tensor{\bar{\EuScript{E}}}{_t_t} = 8 \pi r^3 \tensor{\tilde{\tau}}{_t} . \label{eq:mGravEFEtt}
\end{align}
Similarly, Eqs.~\eqref{eq:mEFEtr} and \eqref{eq:mEFErr} can be written explicitly as
\begin{alignat}{2}
	& \partial_t \Sigma + r^2 \tensor{\bar{\EuScript{E}}}{_t^r} = 8 \pi r^2 e^{\bar{h}} ( \Omega \tensor{\bar{\tau}}{_t^r} + \tensor{\tilde{\tau}}{_t^r}) , \label{eq:mGravEFEtr} \\
	\begin{split}
	& \Sigma \partial_r \bar{C} -  ( r - \bar{C}) (4\Sigma \partial_r \bar{h} + \partial_r \Sigma) \\
	& \qquad \hspace{1.65mm} + 2( r - \bar{C})^2 \partial_r \Omega + r^3 \tensor{\bar{\EuScript{E}}}{^r^r} = 8 \pi r^3 \tensor{\tilde{\tau}}{^r} . \label{eq:mGravEFErr}
	\end{split}
\end{alignat}

\section{Self-consistent solutions in GR} \label{sec:GR}

Here we give a brief summary of the relevant properties of the self-consistent solutions in GR \cite{bmmt:19,mt:20,t:19}. In accord with the previous section (and in anticipation of the notation we use in Sec.~\ref{sec:MTG}), we label functions of pure classical GR (i.e.\ $\lambda=0$) with a bar, e.g.\ the metric functions $\bar{C}$ and $\bar{h}$. The Einstein field equations for $\tensor{\bar{G}}{_t_t}$, $\tensor{\bar{G}}{_t^r}$, and $\tensor{\bar{G}}{^r^r}$ are expressed in terms of the metric functions $\bar{C}$ and $\bar{h}$ as follows:
\begin{align}
	\partial_r \bar{C} &= 8 \pi r^2 \tensor{\bar{\tau}}{_t} / \bar{f} , \label{eq:EFEGRtt}\\
	\partial_t \bar{C} &= 8 \pi r^2 e^{\bar{h}} \tensor{\bar{\tau}}{_t^r} , \label{eq:EFEGRtr} \\
	\partial_r \bar{h} &= 4 \pi r \left( \tensor{\bar{\tau}}{_t} + \tensor{\bar{\tau}}{^r} \right) / \bar{f}^2 . \label{eq:EFEGRrr}
\end{align}

Only two distinct classes of dynamic solutions are possible \cite{t:19}. With respect to the regularity conditions of Eqs.~\eqref{regT} and \eqref{regT2}, they correspond to the values $k= 0$ and $k= 1$.

\subsection{$\boldsymbol{k = 0}$ class of solutions}

In the $k=0$ class of solutions, the limiting form of the reduced EMT components is given by
\begin{align}
	\tensor{\bar{\tau}}{_t} \to - \bar{\Upsilon}^2(t) \; \; , \; \; \tensor{\bar{\tau}}{^r} \to - \bar{\Upsilon}^2(t) \; \; , \; \; \tensor{\bar{\tau}}{_t^r} \to \pm \bar{\Upsilon}^2(t) \; , \label{tau0}
\end{align}
for some function $\bar{\Upsilon}(t)$. The leading terms of the metric functions are
\begin{align}
   \bar{C} &= r_\sg - 4 \pi r_\sg^{3/2} \bar{\Upsilon} \sqrt{x} + \cO(x) , \\
   \bar{h} &= - \frac{1}{2} \ln \frac{x}{\bar{\xi}} + \cO(\sqrt{x}) ,
\end{align}
where $\bar{\xi}(t)$ is determined by the asymptotic properties of the solution. Higher-order terms depend on the higher-order terms in the EMT expansion and will be discussed in Sec.~\ref{sec:MTG}. Consistency of the Einstein equations implies
\begin{align}
	r_\sg' = \pm 4 \bar{\Upsilon} \sqrt{\pi r_\sg \bar{\xi}} . \label{eq:GRcCond}
\end{align}
The null energy condition requires $\tensor{T}{_\mu_\nu} \tensor{l}{^\mu} \tensor{l}{^\nu} \geqslant 0$ for all null vectors $\tensor{l}{^\mu}$ \cite{wald:84,ks:20}. It is violated by radial vectors $\tensor{l}{^{\hat{a}}} = (1, \mp 1,0,0)$ for both the evaporating and accreting solutions, respectively.

The accreting solution $r_\sg'(t)>0$ leads to a firewall: energy density, pressure and flux experienced by an infalling observer diverge at the apparent horizon \cite{t:19}. The resulting averaged negative energy density in the reference frame of a geodesic observer violates a particular  quantum energy inequality \cite{ks:20,ko:15}. Unless we accept that semiclassical physics breaks down already at the horizon scale, this contradiction implies that a PBH cannot grow after its formation \cite{t:19}. Hence we consider only evaporating $r_\sg'(t)<0$ PBHs in what follows.

Matching our results with the standard semiclassical results on black hole evaporation (and accepting that the metric is sufficiently close to the ingoing Vaidya metric with decreasing mass, see \cite{bmt:19} for details) results in
\begin{align}
	\bar{\xi} \sim \frac{\alpha}{r_\sg} ,
\end{align}
where the black hole evaporates according to $r_\sg'(t) = - \alpha / r_\sg^2$ \cite{wald:84,fn:book}. Outside of the apparent horizon the geometry differs from the Schwarzschild metric at least on the scale $r - r_\sg \eqdef x \sim \bar{\xi}$.

\subsection{$\boldsymbol{k = 1}$ solution}

In the second class of solutions $k=1$ and the limiting form of the EMT expansion is given by functions $\tensor{\bar{\tau}}{_a} \propto \bar{f}$. Again, accretion leads to a firewall and thus we will consider only evaporating solutions. It has been shown that dynamic solutions are consistent only in a single case \cite{mt:20}, where in the Schwarzschild frame the energy density $\rho(r_\sg) = \bar{E}$ and pressure $p(r_\sg) = \bar{P}$ at the apparent horizon are given by
\begin{align}
	\bar{E} = - \bar{P} = 1/(8\pi r_\sg^2) .
\end{align}
Since this is their maximal possible value this $k=1$ solution is referred to as extreme \cite{mt:20}. The $k=1$ metric functions are
\begin{align}
  \bar{C} &= r - c_{32} x^{3/2} + \cO(x^2) , \label{eq:GRk1C} \\
  \bar{h} &= - \frac{3}{2} \ln \frac{x}{\bar{\xi}} + \cO(\sqrt{x}) . \label{eq:GRk1h}
\end{align}
Consistency of the Einstein equations then implies
\begin{align}
	r_\sg' = -  c_{32} \bar{\xi}^{3/2} / r_\sg . \label{k1pder}
\end{align}
For future reference we note here that for the  $k=1$ solution the Ricci scalar is given by
\begin{align}
	\bar{R} = 2/r_\sg^2 + \mathcal{O}(x) . \label{eq:k1Ricci}
\end{align}

Evaporating black holes are conveniently represented in $(v,r)$ coordinates, and the limiting form of the $k=0$ solution as $r \to r_\sg$ is a Vaidya metric with decreasing Misner\textendash{}Sharp mass $C_+(v)' < 0$ \cite{bmt:19}. Using $(v,r)$ coordinates to describe geometry at the formation of the first marginally trapped surface reveals how the two classes of solutions are connected (see Ref.~\cite{mt:20} for details): at its formation, a PBH is described by a $k=1$ solution with $\bar{E} = - \bar{P} = 1/(8\pi r_\sg^2)$. It immediately switches to the $k=0$ solution. However, the abrupt transition from $f^1$ to $f^0$ behavior does not lead to discontinuities in the curvature scalars or other physical quantities that could potentially be measured by a local or quasilocal observer.

\section{Self-consistent solutions in MTG}		\label{sec:MTG}
To describe perturbative PBH solutions in MTG the equations must satisfy the same consistency relations as their GR counterparts. Taking the GR solutions as the zeroth-order approximation, we express the functions describing the MTG metric $\sg_\lambda = \bar{\sg} + \lambda \tilde{\sg}$ and thus represent the modified Einstein equations as series in integer and half-integer powers of $x \defeq r - r_\sg$. Their order-by-order solution results in formal expressions for $\Sigma(t,r)$ and $\Omega(t,r)$. However, we also obtain a number of consistency conditions that must be satisfied identically in order for a given theory to admit formation of a PBH. The GR solutions with $k \in \lbrace 0,1 \rbrace$ are sufficiently different to merit a separate treatment provided in Subsec.~\ref{subsec:k0BHs} and \ref{subsec:k1BHs}, respectively.

In both instances, power expansions in various expression have to match up to allow for self-consistent solutions of the modified Einstein equations. Moreover, the relations between the EMT components that are given by Eqs.~\eqref{thev}--\eqref{ther} must hold separately for both the unperturbed terms and the perturbations.

For a given MTG (that is defined by the set of parameters $\lbrace a_1, a_2, a_3, \cdots \rbrace$ in Eq.~\eqref{eq:eftL}) these constraints may conceivably lead to several outcomes: first, it is possible that some of the terms in the Lagrangian Eq.~\eqref{eq:eftL} contribute terms to $\tensor{\bar{\eE}}{_\mu_\nu}$ such that their expansions around $x=0$ lead to terms that diverge stronger than any other terms in Eqs.~\eqref{eq:mGravEFEtt}--\eqref{eq:mGravEFErr}. If only one higher-order curvature term is responsible for such behavior, then such a theory cannot produce perturbative PBH solutions, and only nonperturbative solutions may be possible or the corresponding coefficient $a_i \equiv 0$. If the divergences originate from several terms, they can either cancel if a particular relationship exists between their coefficients $a_i, a_{i'},\ldots$, or not. In the former case the existence of perturbative PBH solutions imposes a constraint, not on the form of the available terms, but on the relationships between their coefficients.

It is also possible that, as it happens in the Starobinsky model (Sec.~\ref{sec:Sgen}), divergences of the terms $\tensor{\bar{\eE}}{_\mu_\nu}$ match the divergences of the GR terms. The constraints can then be satisfied (i) identically (providing us with no additional information); (ii) only for a particular combination of the coefficients $a_i$, thereby constraining the possible classes of MTG; (iii) only in the presence of particular higher-order terms, irrespective of the coefficients, and only for certain unperturbed solutions. In the last scenario, where only certain GR solutions are consistent with a small perturbation, this should be interpreted as an argument against the presence of that particular term in the Lagrangian of Eq.~\eqref{eq:eftL}.

There is \textit{a priori} no reason why $\bar \sg_{\mu\nu} \gg\lambda \tilde \sg_{\mu\nu}$ should hold in some boundary layer around $r_\sg$ \cite{mvc:20,bc-18}. If this condition is not satisfied, then the classification scheme of the GR solutions and a mandatory violation of the null energy condition are not necessarily true. We discuss some of the properties of the solutions without a GR limit and derive the necessary conditions for their existence in Sec.~\ref{sol:lam}.

Throughout this section we use the letter
$j \in \mathbb{Z} \frac{1}{2}$
to label integer and half-integer coefficients and powers of $x$ in series expansions and $\ell$ to refer to generic coefficients. Since we give explicit expressions only for the first few terms in each expression, we write $c_{12}$ instead of $c_{1/2}$, $h_{12}$ instead of $h_{1/2}$, and similarly for higher orders and coefficients of the EMT expansion.

\subsection{Black holes of the $\boldsymbol{k=0}$ type}  \label{subsec:k0BHs}

For the $k=0$ class of solutions the leading terms in the metric functions of classical GR are given as series in powers of $x \defeq r - r_\sg$ as
\begin{align}
	\begin{split}
		\bar{C} &= r_\sg - c_{12} \sqrt{x} + \sum\limits_{1 \leqslant j \in \mathbb{Z}\frac{1}{2}}^\infty c_j x^j \\
		&= r_\sg - c_{12} \sqrt{x} + c_1 x + \mathcal{O}(x^{3/2}) , \label{eq:k0_metric_C}
	\end{split} \\
	\begin{split}
		\bar{h} &= - \frac{1}{2} \ln \frac{x}{\bar\xi} + \sum\limits_{\frac{1}{2} \leqslant j \in \mathbb{Z}\frac{1}{2}}^\infty h_j x^j \\
		&= - \frac{1}{2} \ln \frac{x}{\bar\xi} + h_{12} \sqrt{x} + \mathcal{O}(x) , \label{eq:k0_metric_h}
	\end{split}
\end{align}
where
\begin{align}
	c_{12} &= 4 \sqrt{\pi} r_\sg^{3/2} \bar\Upsilon, \quad c_1 = \frac{1}{3} + \frac{4 \sqrt{\pi} r_\sg^{3/2} (\tensor{\bar{\tau}}{_t})_{12}}{3 \bar{\Upsilon}}, \\
	h_{12} &= \frac{2 \bar{\Upsilon} + \sqrt{\pi} r_\sg^{3/2} \left( 3 (\tensor{\bar{\tau}}{^r})_{12} - (\tensor{\bar{\tau}}{_t})_{12} \right)}{6 \sqrt{\pi} r_\sg^{3/2} \bar{\Upsilon}^2},
\end{align}
and higher-order coefficients of the metric functions are related to higher-order terms in the EMT expansion
\begin{align}
	\tensor{\bar{\tau}}{_a} = -\bar\Upsilon^2 + \sum\limits_{\frac{1}{2} \leqslant j \in \mathbb{Z}\frac{1}{2}}^\infty (\tensor{\bar{\tau}}{_a})_j x^j ,
\end{align}
where $a \in \lbrace \tensor{}{_t} , \tensor{}{_t^r}, \tensor{}{^r} \rbrace \equiv \lbrace \tensor{}{_t_t}, \tensor{}{_t^r}, \tensor{}{^r^r} \rbrace$. We omit the explicit specification $j \in \mathbb{Z}\frac{1}{2}$ from the summation range in what follows.

Regularity of the metric at the apparent horizon and consistency of the Einstein equations establish
algebraic and differential relations between various coefficients. In particular, using Eqs.~\eqref{thev}--\eqref{ther} and Eq.~\eqref{eq:EFEGRtr}, we find
\begin{align}
	(\tensor{\bar{\tau}}{_t})_{12} + (\tensor{\bar{\tau}}{^r})_{12} = 2 (\tensor{\bar{\tau}}{_t^r})_{12} ,
\end{align}
and
\begin{align}
	r_\sg' = - 4 \bar{\Upsilon} \sqrt{\pi r_\sg \bar{\xi}} . \label{con=0-1}
\end{align}

The expansion of $e^{2h}$ that is given by Eq.~\eqref{hdiver2} is obtained as follows: separating the logarithmically divergent part of $h(t,x)$ from the rest, Eq.~\eqref{hdiver} allows to write
\begin{align}
	e^{2h}=\frac{\bar\xi+\lambda\tilde\xi}{x}e^{2\bar \chi+2\lambda \omega}
\end{align}
for some $\tilde\xi(t)$, where $\bar{\chi} = \sum_j h_j x^j$ and $\omega = \sum_j \omega_j x^j$
are convergent functions. First-order expansion in $\lambda$ then leads to Eq.~\eqref{hdiver2} with
\begin{align}
	\Omega=\frac{\tilde\xi}{2\bar \xi} + \omega .
\end{align}
Therefore, the first-order corrections of Eqs.~\eqref{eq:pcSigma}--\eqref{eq:pcOmega} to the metric functions of Eqs.~\eqref{eq:k0_metric_C}--\eqref{eq:k0_metric_h} are given by the series
\begin{align}
	\Sigma &= \sumj \sigma_j x^j = \sigma_{12} x^{1/2} + \sigma_1 x  + \cO(x^{3/2}) , \label{eq:k0_correction_Sigma} \\
	\Omega &= \frac{\tilde{\xi}}{2 \bar{\xi}} + \sumj \omega_j x^j = \frac{\tilde{\xi}}{2 \bar{\xi}} + \omega_{12} x^{1/2} + \cO(x) . \label{eq:k0_correction_Omega}
\end{align}
These two functions can be expressed in terms of the unperturbed solution and corrections $\tensor{\tilde{\tau}}{_a}$ to the EMT from the series expansion of Eqs.~\eqref{eq:mGravEFEtt}--\eqref{eq:mGravEFErr}.
These equations contain various divergent expressions. For example, the term $\Sigma\pad_r\bar C$ as well as all other terms apart from $e^{-2\bar h} \tensor{\bar{\EuScript{E}}}{_t_t}$ in Eq.~\eqref{eq:mGravEFEtt} are finite when $x \to 0$. Then Eq.~\eqref{eq:k0_metric_h} implies that the series expansion of  $\tensor{\bar{\EuScript{E}}}{_t_t}$ starts with a term that is proportional to $1/x$. Performing the same analysis for the two remaining Einstein equations Eqs.~\eqref{eq:mGravEFEtr}--\eqref{eq:mGravEFErr} yields the decompositions
\begin{align}
	\tensor{\bar{\EuScript{E}}}{_t_t} &= \frac{\ae_{\bar{1}}}{x} + \frac{\ae_{\overbar{12}}}{\sqrt{x}} + \ae_0 x^0 + \sumj \ae_j x^j, \label{eq:mEFEk0tt} \\
	\tensor{\bar{\EuScript{E}}}{_t^r} &= \frac{{\oe}_{\overbar{12}}}{\sqrt{x}} + \oe_0 x^0 + \sumj \oe_j x^j , \label{eq:mEFEk0tr} \\
	\tensor{\bar{\EuScript{E}}}{^r^r} &= \o_0 + \sumj \o_j x^j , \label{eq:mEFEk0rr}
\end{align}
of the modified gravity terms that should hold for any $\eF(\tensor{\sg}{_\mu_\nu}, \tensor{R}{_\mu_\nu_\rho_\sigma})$, where indices of coefficients of negative exponents of $x$ are labeled by a bar.

From the requirement that the Ricci scalar $R[g_{\lambda}]$ be finite at the horizon, we obtain the condition
\begin{align}
	\sigma_{12 \vert R} = \frac{\tilde{\xi}}{2 \bar{\xi}} c_{12} = \frac{2 \sqrt{\pi} r_\sg^{3/2} \tilde{\xi} \bar{\Upsilon}}{\bar{\xi}} .
\end{align}
We use additional subscripts (e.g.\ \enquote{$|R$} in the expression above) to indicate what equation was used to derive the explicit expression.

The perturbative contributions Eqs.~\eqref{eq:k0_correction_Sigma}--\eqref{eq:k0_correction_Omega} to the metric functions Eqs.~\eqref{eq:k0_metric_C}--\eqref{eq:k0_metric_h} are obtained order-by-order from the series solutions of Eqs.~\eqref{eq:mGravEFEtt}--\eqref{eq:mGravEFErr}. Expressions for every expansion coefficient can be obtained separately from each equation. Matching of the expressions then allows to identify the coefficients $\ae_\ell$, $\oe_\ell$, $\o_\ell$ of the modified gravity terms Eqs.~\eqref{eq:mEFEk0tt}--\eqref{eq:mEFEk0rr}.
Expressions for $\sigma_{12}$ for instance are obtained from the lowest-order coefficients of Eqs.~\eqref{eq:mGravEFEtt}--\eqref{eq:mGravEFErr}. As a result, we obtain three independent constraints
\begin{align}
	\sigma_{12|R} = \sigma_{12|tt} = \sigma_{12|tr} = \sigma_{12|rr} . \label{eq:sigma12comp1}
\end{align}
They are simultaneously satisfied (see Appendix \ref{app:subsec:k0pertCoCoeff}) if
\begin{align}
	\ae_{\bar{1}} &= - 8 \pi \tilde{\xi} \bar{\Upsilon}^2 \; , \;
	\oe_{\overbar{12}} = - \frac{8 \pi \tilde{\xi} \bar{\Upsilon}^2}{\sqrt{\bar{\xi}}} \; , \;
	\o_0 = - \frac{8 \pi \tilde{\xi} \bar{\Upsilon}^2}{\bar{\xi}} \; .
	\label{eq:k0AE1oe12o0}
\end{align}
These three equations not only identify the function $\tilde \xi(t)$ in terms of unperturbed quantities, but also establish the two relations
\begin{align}
	\ae_{\bar{1}} &= \sqrt{\bar{\xi}} \oe_{\overbar{12}} = \bar{\xi} \o_0		\label{con0-1}
\end{align}
between the leading expansion coefficients of the MTG terms.
Similarly, the next-highest order coefficients of Eqs.\ \eqref{eq:mGravEFEtt}\textendash{}\eqref{eq:mGravEFErr} allow to obtain expressions for $\sigma_1$, see Appendix \ref{app:subsec:k0pertCoCoeff}. Comparison of
\begin{align}
	\sigma_{1 \vert tr}(\omega_{12}) = \sigma_{1 \vert rr}(\omega_{12}) \label{eq:k1sigma1trrr}
\end{align}
gives an expression for $\omega_{12}$. Substitution into Eq.~\eqref{eq:k1sigma1trrr} and subsequent comparison of $\sigma_{1 \vert tr} = \sigma_{1 \vert rr}$ with $\sigma_{1 \vert tt}$ gives a relation between the next-highest order coefficients, namely
\begin{align}
	\ae_{\overbar{12}} = 2 \sqrt{\bar{\xi}} \oe_0 - \bar{\xi} \o_{12} .     \label{con0-2}
\end{align}
The modified gravity terms $\tensor{\bar{\eE}}{_\mu_\nu}$ must adhere to the expansion structures in Eqs.~\eqref{eq:mEFEk0tt}--\eqref{eq:mEFEk0rr}, and the relations Eqs.~\eqref{con0-1} and \eqref{con0-2} between their coefficients must be satisfied identically. Otherwise the MTG solutions do not exist. For terms in the metric functions of order $\mathcal{O}(x^{3/2})$ and higher the three equations Eqs.~\eqref{eq:mGravEFEtt}--\eqref{eq:mGravEFErr} for the $tt$, $tr$, and $rr$ component contain three additional independent variables $(\tensor{\bar{\tau}}{_a})_{j \geqslant 3/2}$ and will therefore not lead to any additional constraints.

It is worth pointing out that the analogs of Eqs.~\eqref{con0-1} and \eqref{con0-2} are also satisfied by the coefficients of the corresponding metric tensor and Ricci tensor components themselves, i.e.\
\begin{align}
		(\tensor{\bar{\sg}}{_t_t})_{\bar{1}} &= \sqrt{\bar{\xi}} (\tensor{\bar{\sg}}{_t^r})_{\overbar{12}} = \bar{\xi} (\tensor{\bar{\sg}}{^r^r})_0 = 0 , \label{eq:k0con1metric} \\
		(\tensor{\bar{R}}{_t_t})_{\bar{1}} &= \sqrt{\bar{\xi}} (\tensor{\bar{R}}{_t^r})_{\overbar{12}} = \bar{\xi} (\tensor{\bar{R}}{^r^r})_0 = - c_{12}^2 \bar{\xi} / (2 r_\sg^3) , \label{eq:k0con1Ricci}
\end{align}
and
\begin{align}
	(\tensor{\bar{\sg}}{_t_t})_{\overbar{12}} &= 2 \sqrt{\bar{\xi}} (\tensor{\bar{\sg}}{_t^r})_0 - \bar{\xi} (\tensor{\bar{\sg}}{^r^r})_{12} = - c_{12} \bar{\xi} / r_\sg , \label{eq:k0con2metric} \\
	(\tensor{\bar{R}}{_t_t})_{\overbar{12}} &= 2 \sqrt{\bar{\xi}} (\tensor{\bar{R}}{_t^r})_0 - \bar{\xi} (\tensor{\bar{R}}{^r^r})_{12} , \label{eq:k0con2Ricci}
\end{align}
where Eq.~\eqref{eq:k0con1metric} is satisfied trivially and Eq.~\eqref{eq:k0con2metric} simplifies to $(\tensor{\bar{\sg}}{_t_t})_{\overbar{12}} = - \bar{\xi} (\tensor{\bar{\sg}}{^r^r})_{12}$ due to the diagonal form of the metric tensor $\tensor{\bar{\sg}}{_t^r} = 0$ (see Eq.~\eqref{eq:metric}). Explicit expressions for the coefficients of the Ricci tensor components in Eq.~\eqref{eq:k0con2Ricci} are provided in Appendix~\ref{app:k0expressions}, see Eqs.~\eqref{eq:k0Rttb12}--\eqref{eq:k0Rrr12}.

\subsection{Black holes of the $\boldsymbol{k=1}$ type}  \label{subsec:k1BHs}

The EMT expansion for the $k = 1$ solution is given in terms of $x \defeq r - r_\sg$ by
\begin{align}
	\tensor{\tau}{_t} &= \tensor{\bar{\tau}}{_t} + \lambda \tensor{\tilde{\tau}}{_t} = \bar{f} \left( \bar{E} + \lambda \tilde{E} \right) + \sum\limits_{j \geqslant 2} e_j x^j , \label{eq:k1taut} \\
	\tensor{\tau}{_t^r} &= \tensor{\bar{\tau}}{_t^r} + \lambda \tensor{\tilde{\tau}}{_t^r} = \bar{f} \left( \bar{\Phi} + \lambda \tilde{\Phi} \right) + \sum\limits_{j \geqslant 2} \phi_j x^j , \label{eq:k1tautr} \\
	\tensor{\tau}{^r} &= \tensor{\bar{\tau}}{^r} + \lambda \tensor{\tilde{\tau}}{^r} = \bar{f} \left( \bar{P} + \lambda \tilde{P} \right) + \sum\limits_{j \geqslant 2} p_j x^j , \label{eq:k1taur}
\end{align}
where $\bar{E} = - \bar{P} = 1/(8 \pi r_\sg^2)$ and $\bar{\Phi} = 0$. To improve readability and clarify the connection to physical quantities (energy, pressure, flux) we set $(\tensor{\bar{\tau}}{_t})_ j \eqdef \bar{e}_j$, $(\tensor{\bar{\tau}}{^r})_ j \eqdef \bar{p}_j$, and $(\tensor{\bar{\tau}}{_t^r})_ j \eqdef \bar{\phi}_j$, and analogously for the perturbative coefficients $(\tensor{\tilde{\tau}}{_a})_ j$. Additional relations between the coefficients are obtained from Eqs.~\eqref{thev}--\eqref{thevr}, i.e.\
\begin{align}
	& \tilde{E} + \tilde{P} = 2 \tilde{\Phi} , \label{eq:k1TildeTauRel32} \\
	& \bar{e}_2 = \bar{p}_2 = \bar{\phi}_2 , \quad \bar{e}_{52} + \bar{p}_{52} = 2 \bar{\phi}_{52} , \\
	& \tilde{e}_2 + \tilde{p}_2 =2\tilde{\phi}_2 , \quad \tilde{e}_{52} + \tilde{p}_{52} = 2 \tilde{\phi}_{52} , \label{eq:k1TildeTauRel}
\end{align}
for the two next-highest orders $j = 2, \frac{5}{2}$.
Recall that (cf.\ Eqs.~\eqref{eq:GRk1C}--\eqref{eq:GRk1h}) the leading terms in the metric functions of classical GR are given as series in powers of $x \defeq r - r_\sg$ as
\begin{align}
	\bar{C} &= r_\sg + x - c_{32} x^{3/2} + \mathcal{O}(x^2) , \label{eq:k1_metric_C} \\
	\bar{h} &= - \frac{3}{2} \ln \frac{x}{\bar{\xi}} + h_{12} \sqrt{x} + \mathcal{O}(x) , \label{eq:k1_metric_h}
\end{align}
with coefficients
\begin{align}
	c_{32} &= 4 r_\sg^{3/2} \sqrt{- \pi \bar{e}_2 / 3} , \label{eq:k1c32} \\
	h_{12} &= \frac{3}{14 \bar{e}_2} \left( \frac{4 \sqrt{- 3 \bar{e}_2 / \pi}}{r_\sg^{5/2}} + 5 \bar{e}_{52} - 7 \bar{p}_{52} \right) . \label{eq:k1h12}
\end{align}
Higher-order coefficients are obtained from higher-order terms of the EMT expansion using consistency of the Einstein equations, e.g.\ the next-highest order coefficient of Eq.~\eqref{eq:k1_metric_C} is
\begin{align}
	c_2 &= \frac{4}{7 r_\sg} \left( 1 + \frac{r_\sg^{5/2} \sqrt{3 \pi} \bar{e}_{52}}{\sqrt{- \bar{e}_2}} \right) .
\end{align}
In addition, consistency of the Einstein equations requires Eq.~\eqref{k1pder} and
\begin{align}
	\bar{p}_{52} = \frac{2 \sqrt{- \bar{e}_2}}{\sqrt{3 \pi} r_\sg^{5/2}} + \bar{e}_{52} .
	\label{eq:k1p52}
\end{align}
Substituting Eq.~\eqref{eq:k1p52} into Eq.~\eqref{eq:k1h12} we obtain the identity
\begin{align}
	c_2 = c_{32} h_{12} , \label{eq:k1c2c32h12}
\end{align}
which leads to many simplifying cancellations, e.g.\ the absence of the $\sqrt{x}$ term in the Ricci scalar $\bar{R}$ (cf.\ Eq.~\eqref{eq:k1Ricci}) due to $R_{12} \propto c_{32} h_{12} - c_2$, where $R_{12}$ denotes the $\sqrt{x}$ coefficient of $\bar{R}$.

Again, the expansion of $e^{2h}$ that is given by Eq.~\eqref{hdiver2} is obtained by separating the logarithmic part of $h(t,x)$ from the rest. From the expansion
\begin{align}
	e^{2h} = \left( \frac{\bar\xi+\lambda\tilde\xi}{x}\right)^3 e^{2\bar \chi+2\lambda \omega},
\end{align}
we then obtain Eq.~\eqref{hdiver2} with
\begin{align}
	\Omega = \frac{3 \tilde{\xi}}{2 \bar{\xi}} + \omega.
\end{align}
The series expansions of the perturbative corrections of Eqs.~\eqref{eq:pcSigma}--\eqref{eq:pcOmega} are therefore given by the power series
\begin{align}
	\Sigma &= \sum\limits_{j \geqslant \frac{3}{2}}^\infty \sigma_j x^j = \sigma_{32} x^{3/2} + \sigma_2 x^2 + \mathcal{O}(x^{5/2}) , \label{eq:k1_correction_Sigma} \\
	\Omega &= \frac{3 \tilde{\xi}}{2 \bar{\xi}} + \sumj \omega_j x^j = \frac{3 \tilde{\xi}}{2 \bar{\xi}} + \omega_{12} \sqrt{x} + \omega_1 x + \mathcal{O}(x^{3/2}) . \label{eq:k1_correction_Omega}
\end{align}
Finiteness of the Ricci scalar at the horizon requires Eq.~\eqref{k1pder} and
\begin{align}
	\sigma_{32 \vert R} = \frac{3 \tilde{\xi}}{2 \bar{\xi}} c_{32} 
=\frac{2 r_\sg^{3/2} \tilde{\xi} \sqrt{- 3 \pi \bar{e}_2}}{\bar{\xi}} . \label{eq:k1sigma32R}
\end{align}
The expansion structure of the modified gravity terms $\tensor{\bar{\EuScript{E}}}{_\mu_\nu}$ is obtained analogous to Sec.~\ref{subsec:k0BHs}. We find
\begin{align}
	\tensor{\bar{\EuScript{E}}}{_t_t} &= \frac{\ae_{\overbar{32}}}{x^{3/2}} + \frac{\ae_{\bar{1}}}{x} + \frac{\ae_{\overbar{12}}}{\sqrt{x}} + \ae_0 + \sumj \ae_j x^j , \label{eq:k1Ettstructure} \\
	\tensor{\bar{\EuScript{E}}}{_t^r} &= \oe_0 + \sumj \oe_j x^j , \label{eq:k1Etrstructure} \\	
	\tensor{\bar{\EuScript{E}}}{^r^r} &= \sum\limits_{j \geqslant \frac{3}{2}}^\infty \o_j x^j . \label{eq:k1Errstructure}
\end{align}
The equation for the $x^0$ coefficient of the $tr$ component Eq.~\eqref{eq:mGravEFEtr} allows to identify
\begin{align}
	\tilde{E}_{\vert tr} &= \frac{\oe_0 r_\sg}{8 \pi \bar{\xi}^{3/2} c_{32}} . \label{eq:k1TildeE}
\end{align}
Substitution of Eq.~\eqref{eq:k1TildeE} into the expression $\sigma_{32 \vert tt}(\tilde{E})$ obtained from Eq.~\eqref{eq:mGravEFEtt} and subsequent comparison with the expression $\sigma_{32 \vert rr}$ obtained from Eq.~\eqref{eq:mGravEFErr} establishes the relation
\begin{align}
	\ae_{\overbar{32}} &= 2 \bar{\xi}^{3/2} \oe_0 - \bar{\xi}^{3} \o_{32}		\label{eq:k1cond1}
\end{align}
between the lowest-order coefficients of the MTG terms, see Appendix \ref{app:subsec:k1pertCoCoeff}. Similarly, by substituting Eq.~\eqref{eq:k1TildeE} into the expression for $\sigma_{2 \vert tt}$, and $\tilde{\xi}_{\vert tr}$ obtained from the $\sqrt{x}$ coefficient of Eq.~\eqref{eq:mGravEFEtr} into the expression Eq.~\eqref{eq:k1sigma32R} for $\sigma_{32 \vert R}$, we can derive two distinct expressions for the sum $\tilde{e}_2 + \tilde{p}_2$ by comparison of $\sigma_{2 \vert tt}$ and $\sigma_{2 \vert rr}$ obtained from Eqs.~\eqref{eq:mGravEFEtt} and \eqref{eq:mGravEFErr}, respectively, as well as comparison of $\sigma_{32 \vert R}$ and $\sigma_{32 \vert tt}$. Their identification establishes the additional relation
\begin{align}
		\ae_{\bar{1}} &= 2 \bar{\xi}^{3/2} \left( h_{12} \oe_0 + \oe_{12} \right)	 - \bar{\xi}^3 \left( 2 h_{12} \o_{32} + \o_2 \right) \label{eq:k1cond2}
\end{align}
between the modified gravity coefficients of Eqs.~\eqref{eq:k1Ettstructure}\textendash{}\eqref{eq:k1Errstructure}. A detailed derivation with explicit expressions is provided in Appendix \ref{app:subsec:k1pertCoCoeff}. Analogous to the class of $k = 0$ black hole solutions discussed in Sec.~\ref{subsec:k0BHs}, the modified gravity terms $\tensor{\bar{\eE}}{_\mu_\nu}$ of any self-consistent MTG must follow the expansion structures prescribed by Eqs.~\eqref{eq:k1Ettstructure}--\eqref{eq:k1Errstructure} and identically satisfy the two relations Eqs.~\eqref{eq:k1cond1}--\eqref{eq:k1cond2} to be compatible with black hole solutions of the $k=1$ type. Again, consideration of higher-order coefficients in Eqs.~\eqref{eq:mGravEFEtt}-\eqref{eq:mGravEFErr} introduces new independent variables and will thus not yield any additional constraints.

Once more, the analogs of the MTG coefficient relations Eqs.~\eqref{eq:k1cond1}--\eqref{eq:k1cond2} are also satisfied by the coefficients of the corresponding metric tensor and Ricci tensor components themselves, i.e.\
\begin{align}
	(\tensor{\bar{\sg}}{_t_t})_{\overbar{32}} &= - \bar{\xi}^3 (\tensor{\bar{\sg}}{^r^r})_{32} = - c_{32} \bar{\xi}^3 / r_\sg , \label{eq:k1con1metric} \\
	(\tensor{\bar{R}}{_t_t})_{\overbar{32}} &= 2 \bar{\xi}^{3/2} (\tensor{\bar{R}}{_t^r})_0 - \bar{\xi}^3 (\tensor{\bar{R}}{^r^r})_{32} = 0 , \label{eq:k1con1Ricci}
\end{align}
and
\begin{align}
	\begin{split}
		(\tensor{\bar{\sg}}{_t_t})_{\bar{1}} &= - \bar{\xi}^3 \bigl( 2 h_{12} (\tensor{\bar{\sg}}{^r^r})_{32} + (\tensor{\bar{\sg}}{^r^r})_{2} \bigr) \\
		&= - c_{32} h_{12} \bar{\xi}^3 / r_\sg ,
		\label{eq:k1con2metric}
	\end{split}
	\\
	\begin{split}
		(\tensor{\bar{R}}{_t_t})_{\bar{1}} &= 2 \bar{\xi}^{3/2} \left( h_{12} (\tensor{\bar{R}}{_t^r})_0 + (\tensor{\bar{R}}{_t^r})_{12} \right) \\
		& \qquad - \bar{\xi}^3 \bigl( 2 h_{12} (\tensor{\bar{R}}{^r^r})_{32} + (\tensor{\bar{R}}{^r^r})_{2} \bigr) \\
		&= - 3 c_{32}^2 \bar{\xi}^3 / \bigl( 2 r_\sg^3 \bigr) ,
		\label{eq:k1con2Ricci}
	\end{split}
\end{align}
where Eq.~\eqref{eq:k1c2c32h12} was used to simplify the expressions, $(\tensor{\bar{R}}{_t_t})_{\overbar{32}} = (\tensor{\bar{R}}{_t^r})_0 = (\tensor{\bar{R}}{^r^r})_{32} = 0$,
$(\tensor{\bar{R}}{^r^r})_2 = - 3 c_{32}^2 / (2 r_\sg^3)$, and Eqs.~\eqref{eq:k1con1metric} and \eqref{eq:k1con2metric} simplify due to the diagonal form of the metric tensor $\tensor{\bar{\sg}}{_t^r} = 0$ (see Eq.~\eqref{eq:metric}).

\subsection{$\boldsymbol{\lambda}$-expanded solutions} \label{sol:lam}

We now consider solutions where the leading reduced components of the EMT are not dominated by terms of order $\mathcal{O}(\lambda^0)$. To obtain mathematically consistent expressions we have to extend the expansion to terms of order $\mathcal{O}(\lambda^2)$ as higher-order terms, if needed, are obtained analogously.

The $k=0$ solution without GR limit has the following properties: the EMT expansion
\begin{align}
	\begin{aligned}
		\tensor{\tau}{_a} = & \lambda \tilde{\Xi} + \lambda^2 \tilde{\Xi}_{(2)} \\
		& + \sum\limits_{j \geqslant \frac{1}{2}}^\infty \left[ (\tensor{\bar{\tau}}{_a})_j + \lambda (\tensor{\tilde{\tau}}{_a})_j + \lambda^2 \big( \tensor{\tilde{\tau}}{_a^{\hspace*{-1mm} (2)}} \big)_j \right] x^j
	\end{aligned}
	\label{eq:k0tau-non}
\end{align}
corresponds to the case where as $r \to r_\sg$, $\lim \tensor{\tau}{_t} = \lim \tensor{\tau}{^r} = \lim \tensor{\tau}{_t^r}$. The equations below are trivially extendable to the case where the leading term in $\tensor{\tau}{_t^r} = - \lambda \tilde{\Xi}$.

In either case, the metric functions are given by
\begin{align}
	 C &= r_\sg -\lambda \sigma_{12}\sqrt{x} +\sum\limits_{j \geqslant \frac{1}{2}}^\infty \left( \zeta_j + \lambda \sigma_j + \lambda^2 \sigma^{(2)}_j \right) x^j , \label{eq:k0C-non} \\
	 h &= - \frac{1}{2} \ln \frac{x}{\xi} + \sum\limits_{j \geqslant \frac{1}{2}}^\infty \left( \eta_j + \lambda \omega_j + \lambda^2 \omega_j^{(2)} \right) x^j , \label{eq:k0h-non}
\end{align}
similar to the $k= 0$ perturbative solution. Here, the structure of the metric function $h$ was simplified by redefining the time, and the coefficient $c_{12} = \zeta_{12} + \lambda \sigma_{12} + \lambda^2 \sigma^{(2)}_{12}$ was simplified by taking into account the requirement that the Ricci scalar must be finite at the apparent horizon, i.e.\
\begin{align}
	c_{12}\to\lambda\sigma_{12}=-\frac{r_\sg'\, r_\sg}{\sqrt{\xi}}.
\end{align}
Unlike in GR, the sign of $\tilde{\Xi}$ (and $\tilde{\Xi}_{(2)}$) cannot be determined solely from the requirements of existence and consistency of the modified Einstein equations. It is therefore unclear whether or not violation of the null energy condition is a prerequisite for the formation of a PBH. This is in contrast to GR, where such a violation has been shown to be mandatory in a variety of settings \cite{fn:book,he:book,bmmt:19,t:19}.

The expansion structure of the non-GR terms $\tensor{\eE}{_\mu_\nu}$ remains the same as in the perturbative $k=0$ scenario discussed in Subsec.~\ref{subsec:k0BHs}, that is
\begin{align}
	\tensor{\EuScript{E}}{_t_t} &= \frac{\ae_{\bar{1}}}{x} + \frac{\ae_{\overbar{12}}}{\sqrt{x}} + \ae_0 x^0 + \sumj \ae_j x^j, \label{eq:mEFEk0ttlambda} \\
	\tensor{\EuScript{E}}{_t^r} &= \frac{{\oe}_{\overbar{12}}}{\sqrt{x}} + \oe_0 x^0 + \sumj \oe_j x^j , \label{eq:mEFEk0trlambda} \\
	\tensor{\EuScript{E}}{^r^r} &= \o_0 + \sumj \o_j x^j , \label{eq:mEFEk0rrlambda}
\end{align}
where $\ae_\ell \defeq \bar{\ae}_\ell + \lambda \tilde{\ae}_\ell$, and similarly for the coefficients $\oe_\ell$, $\o_\ell$ of the non-GR terms $\tensor{\EuScript{E}}{_t^r}$ and $\tensor{\EuScript{E}}{^r^r}$. Substitution into the generic modified Einstein equations Eqs.~\eqref{eq:mEFEtt}--\eqref{eq:mEFErr} gives
\vspace*{3mm} \\
\begin{tabular*}{\textwidth}{lr}
	\hspace*{-1.75mm}
	\resizebox{.235\linewidth}{!}{$
	\begin{aligned}
		& \mathcal{O}\left( x^0 \right) \\
		& \text{terms of \phantom{Eq.~\eqref{eq:mEFEtt}}} \\
		& \text{Eq.~\eqref{eq:mEFEtt}}
	\end{aligned}
	$}
	\hspace*{-9mm}
	&
	\begin{numcases}{}
		\hspace*{-1.25mm} \vspace{2.5mm} \nonumber \\
		\left( \frac{\bar{\ae}_{\bar{1}}}{\xi} - 8 \pi \tilde{\Xi} \right) \lambda = 0 , \\
		\hspace*{-1.25mm}
		\left( \frac{\tilde{\ae}_{\bar{1}}}{\xi} + 8 \pi \tilde{\Xi}_{(2)} - \frac{\sigma_{12}^2}{2 r_\sg^3} \right) \lambda^2 = 0 , \\
		\hphantom{\left[ \tilde{\oe}_{\overbar{12}} + \frac{\sqrt{\xi}}{2} \left( 16 \pi \tilde{\Xi}_{(2)} - \frac{\sigma_{12}^2}{r_\sg^3} \right) \right] \lambda^2 = 0 ,
		\hspace*{8.65mm}} \nonumber
		\vspace*{-2mm}
	\end{numcases}
	\\
	\hspace*{-1.75mm}
	\resizebox{.235\linewidth}{!}{$
	\begin{aligned}
		& \mathcal{O} \bigl( x^{-1/2} \bigr) \\
		& \text{terms of \phantom{Eq.~\eqref{eq:mEFEtr}}} \\
		& \text{Eq.~\eqref{eq:mEFEtr}}
	\end{aligned}
	$}
	\hspace*{-9mm}
	&
	\begin{numcases}{}
		\hspace*{-1.25mm} \vspace{2.5mm} \nonumber \\
		\left( \bar{\oe}_{\overbar{12}} - 8 \pi \sqrt{\xi} \tilde{\Xi} \right) \lambda = 0 , \\
		\hspace*{-1.25mm}
		\left[ \tilde{\oe}_{\overbar{12}} + \frac{\sqrt{\xi}}{2} \left( 16 \pi \tilde{\Xi}_{(2)} - \frac{\sigma_{12}^2}{r_\sg^3} \right) \right] \lambda^2 = 0 , \\
		\hphantom{\left[ \tilde{\oe}_{\overbar{12}} + \frac{\sqrt{\xi}}{2} \left( 16 \pi \tilde{\Xi}_{(2)} - \frac{\sigma_{12}^2}{r_\sg^3} \right) \right] \lambda^2 = 0 ,
		\hspace*{8.65mm}} \nonumber
		\vspace*{-2mm}
	\end{numcases}
	\\
	\hspace*{-1.75mm}
	\resizebox{.235\linewidth}{!}{$
	\begin{aligned}
		& \mathcal{O} \left( x^0 \right) \\
		& \text{terms of \phantom{Eq.~\eqref{eq:mEFErr}}} \\
		& \text{Eq.~\eqref{eq:mEFErr}}
	\end{aligned}
	$}
	\hspace*{-9mm}
	&
	\begin{numcases}{}
		\hspace*{-1.25mm} \vspace{2.5mm} \nonumber \\
		\left( \bar{\o}_0 - 8 \pi \tilde{\Xi} \right) \lambda = 0 , \\
		\hspace*{-1.25mm}
		\left( \tilde{\o}_0 + 8 \pi \tilde{\Xi}_{(2)} - \frac{\sigma_{12}^2}{2 r_\sg^3} \right) \lambda^2 = 0 , \\
		\hphantom{\left[ \tilde{\oe}_{\overbar{12}} + \frac{\sqrt{\xi}}{2} \left( 16 \pi \tilde{\Xi}_{(2)} - \frac{\sigma_{12}^2}{r_\sg^3} \right) \right] \lambda^2 = 0 ,
		\hspace*{8.65mm}} \nonumber
		\vspace*{-2mm}
	\end{numcases}
	\vspace*{3mm}
\end{tabular*}\\
at the respective leading orders of $x$, and leads to the following constraints: first, the expansion coefficients $(\tensor{\tau}{_a})_{12} = 0 \; \forall a$, where $(\tensor{\tau}{_a})_\ell \defeq (\tensor{\bar{\tau}}{_a})_\ell + \lambda (\tensor{\tilde{\tau}}{_a})_\ell + \lambda^2 (\tensor{\tilde{\tau}}{_a^{\hspace*{-1mm}(2)}})_\ell$, see Eq.~\eqref{eq:k0tau-non}. At the leading expansion order (which is $\mathcal{O}(\lambda)$ since the non-GR terms appear as $\lambda \tensor{\eE}{_\mu_\nu}$ in Eqs.~\eqref{eq:mEFEtt}--\eqref{eq:mEFErr}) the lowest-order $x$ coefficients satisfy
\begin{align}
	\bar{\ae}_{\bar{1}} = \sqrt{\xi} \bar{\oe}_{\overbar{12}} = \xi \bar{\o}_0 = 8 \pi \tilde{\Xi} \xi ,
	\label{con1b}
\end{align}
which is analogous to Eq.~\eqref{con0-1}, but in this case identifies the leading reduced term in the EMT. Similarly, the next-order $\mathcal{O}(\lambda^2)$ expansion coefficients satisfy
\begin{align}
	\tilde{\ae}_{\bar{1}} = \sqrt{\xi} \tilde{\oe}_{\overbar{12}} = \xi \tilde{\o}_0 = - 8 \pi \tilde{\Xi}_{(2)} \xi + \frac{\sigma_{12}^2 \xi}{2 r_\sg^3} .
	\label{con1t}
\end{align}

\section{Black holes in the Starobinsky model} \label{sec:Sgen}

Numerous modifications of GR have been proposed, including theories that involve higher-order curvature invariants. A popular class among these are so-called $\mathfrak{f}(R)$ theories \cite{f-R}, in which the gravitational Lagrangian density $\eL_\mathrm{g}$ is an arbitrary function of the Ricci scalar $R$. In this section, we consider the Starobinsky model \cite{staro} with $\eF=\varsigma R^2$, $\varsigma=16 \pi a_2 / M^2_\mathrm{P}$ (see Eq.~\eqref{eq:eftL}). It is a straightforward extension of GR with quadratic corrections in the Ricci scalar that is of relevance in cosmological contexts. In particular, it is the first self-consistent model of inflation. New horizonless solutions in this model have been identified recently in an analysis \cite{eH.cfS.2020} of static, spherically symmetric, and asymptotically flat vacuum solutions.

\subsection{Modified Einstein equations}

In $\mathfrak{f}(R)$ theories, the relevant equations have a relatively simple form. For the action
\begin{align}
	S=\frac{1}{16\pi}\int \big(\mathfrak{f}(R)+\eL_\mathrm{m}\big)\sqrt{-\sg} \; d^4x + S_\mathrm{b},
\end{align}
where the gravitational Lagrangian $\eL_\mathrm{g} = \mathfrak{f}(R)$, the matter Lagrangian is represented by $\eL_\mathrm{m}$, and $S_\mathrm{b}$ denotes the boundary term, the field equations for the metric $\tensor{\sg}{_\mu_\nu}$ are given by
\begin{align}
	\mathfrak{f}' \tensor{R}{_\mu_\nu} - \frac{1}{2} \mathfrak{f} \tensor{\sg}{_\mu_\nu} + \left( \tensor{\sg}{_\mu_\nu} \square - \nabla_\mu \nabla_\nu \right) \mathfrak{f}' = 8 \pi \tensor{T}{_\mu_\nu} ,
\end{align}
where $\mathfrak{f}' \defeq \partial \mathfrak{f}(R) / \partial R$ and $\square \defeq \tensor{\sg}{^\mu^\nu} \nabla_\mu \nabla_\nu$. It is convenient to set $\mathfrak{f}(R) \eqdef R + \lambda \eF(R)$.
The modified Einstein equations are then
\begin{align} 	
	\tensor{G}{_\mu_\nu} + \lambda \left(\eF' \tensor{R}{_\mu_\nu}
 - \frac{1}{2} \eF \tensor{\sg}{_\mu_\nu}+ \left( \tensor{\sg}{_\mu_\nu} \square - \nabla_\mu \nabla_\nu \right) \eF'\right)
  = 8 \pi \tensor{T}{_\mu_\nu} .
\end{align}
Performing the expansion in $\lambda$ and only keeping terms up to the first order we obtain expressions for the modified gravity terms $\tensor{\bar{\EuScript{E}}}{_\mu_\nu}$, i.e.\
\begin{align}
	\tensor{\bar{\EuScript{E}}}{_\mu_\nu} = \eF^\prime \tensor{\bar{R}}{_\mu_\nu} - \frac{1}{2} \eF\, \tensor{\bar{\sg}}{_\mu_\nu} + \left( \tensor{\bar{\sg}}{_\mu_\nu} \bar{\square} - \bar{\nabla}_\mu \bar{\nabla}_\nu \right) \eF^\prime , \label{eomf}
\end{align}
where all objects labeled by the bar are evaluated with respect to the unperturbed metric $\bar{\sg}$, and $\eF \equiv \eF\big(\bar{R}\big)$.

In spherical symmetry the d'Alembertian is given by
\begin{align}
	\square \eF^\prime &= \left[ \partial_t \partial^t + \partial_r \partial^r + \left( \partial_t  h \right) \partial^t +
 \left( \partial_r h + 2/r \right) \partial^r \right] \eF^\prime . \label{eq:d'AlembertianTerm}
\end{align}
Second-order covariant derivatives of a scalar function can be expressed in terms of partial derivatives, i.e.
\begin{align}	
	\nabla_\mu \nabla_\nu \eF^\prime = \partial_\mu \partial_\nu \eF^\prime - \Gamma_{\mu\nu}^\zeta \partial_\zeta \eF^\prime. \label{eq:sec_cov_der}
\end{align}
In the Starobinsky model $\eF( {R}) = \varsigma  {\bar R}^2 + \mathcal{O}(\lambda)$ and Eqs.~\eqref{eq:mGravEFEtt}--\eqref{eq:mGravEFErr} become
\begin{align}
	\begin{split}
	\tensor{\bar{\EuScript{E}}}{_t_t} / (\lambda \varsigma) &= 2 \bar{R} \tensor{\bar{R}}{_t_t} - \half \bar{R}^2 \tensor{\bar{\sg}}{_t_t} + 2 \big[ \tensor{\bar{\sg}}{_t_t} \big( \pad_t \pad^t + \pad_r \pad^r \\
	& \hspace*{7mm}	
	+ (\pad_t \bar{h}) \pad^t + (\pad_r \bar{h} + 2r^{-1}) \pad^r \big)
	\\
	& \hspace*{7mm}	
	- \pad_t \pad_t + \Gamma^t_{tt} \pad_t + \Gamma^r_{tt} \pad_r \big] \bar{R}
	, \label{eq:fR2Ett}
	\end{split}
	\\
	\tensor{\bar{\EuScript{E}}}{_t^r} / (\lambda \varsigma) &= 2 \bar{R} \tensor{\bar{R}}{_t^r} - 2 \left( \partial_t \partial^r + \Gamma_{tt}^r \partial^t + \Gamma_{tr}^r \partial^r \right) \bar{R} , \label{eq:fR2Etr} \\
	\begin{split}
		\tensor{\bar{\EuScript{E}}}{^r^r} / (\lambda \varsigma) &= 2 \bar{R} \tensor{\bar{R}}{^r^r} - \half \bar{R}^2 \tensor{\bar{\sg}}{^r^r} + 2 \tensor{\bar{\sg}}{^r^r} \big[ \pad_t \pad^t
	\\
	& \hspace*{-7mm}
	+ (\pad_t \bar{h} - \Gamma^r_{r t}) \pad^t
	+ (\pad_r \bar{h} + 2r^{-1} - \Gamma^r_{rr}) \pad^r \big] \bar{R} .
	\label{eq:fR2Err}
	\end{split}
\end{align}

\subsection{Compatibility with the $\boldsymbol{k=0}$ class of black hole solutions} \label{subsec:Compf(R)k0}

With the $k=0$ metric functions Eqs.~\eqref{eq:k0_metric_C}\textendash{}\eqref{eq:k0_metric_h}, the constraint Eq.~\eqref{con=0-1} that is obtained from the requirement that the Ricci scalar be non-divergent leads to cancellations in the Ricci tensor components $\tensor{\bar{R}}{_t_t}$ and $\tensor{\bar{R}}{^r^r}$ which ensures that the MTG terms Eqs.~\eqref{eq:fR2Ett}--\eqref{eq:fR2Err} of the $\tilde{\mathfrak{f}}(\bar{R}) = \varsigma \bar{R}^2$ Starobinsky model conform to the structures of Eqs.~\eqref{eq:mEFEk0tt}--\eqref{eq:mEFEk0rr}. We find that both of the two constraints posed by Eq.~\eqref{con0-1} are satisfied, i.e.\
\begin{align}
	\begin{aligned}
		\ae_{\bar{1}} &= \sqrt{\bar{\xi}} \oe_{\overbar{12}} = \bar{\xi} \o_0 \\
		&= \frac{c_{12}^2 \bar{\xi} \left( -2 \left( R_0 + r_\sg R_1 \right) + h_{12} r_\sg R_{12} \right) - c_{12} r_\sg^2 \sqrt{\bar{\xi}} R_0^\prime}{2 r_\sg^3} ,
	\end{aligned}
	\label{eq:k0_con1_Staro}
\end{align}
where $R_j$ is used to denote coefficients of the Ricci scalar $\bar{R} = \sum_j R_j x^j = R _0 + R _{12} \sqrt{x} + R _1 x + \mathcal{O}(x^{3/2})$. Similarly, the next-highest order coefficients satisfy the constraint of Eq.~\eqref{con0-2}, see App.~\ref{app:k0expressions}, Eqs.~\eqref{eq:k0aeb12}--\eqref{eq:k0o12}.

\subsection{Compatibility with the $\boldsymbol{k=1}$ solution}

Similar to the $k=0$ case, the $k=1$ constraint on the evolution of the horizon radius Eq.~\eqref{k1pder} that is required to ensure consistency of the Einstein equations and finiteness of the Ricci scalar leads to cancellations in $\tensor{\bar{R}}{_t_t}$, and the $k=1$ Starobinsky MTG terms of Eqs.~\eqref{eq:fR2Ett}--\eqref{eq:fR2Err} follow the structures prescribed by Eqs.~\eqref{eq:k1Ettstructure}--\eqref{eq:k1Errstructure}. Using the $k=1$ metric functions Eqs.~\eqref{eq:k1_metric_C}\textendash{}\eqref{eq:k1_metric_h} we obtain the lowest-order coefficients
\begin{align}
	\ae_{\overbar{32}} &= 2 c_{32} \bar{\xi}^3 / r_\sg^5 , \label{eq:k1AE32} \\
	\ae_{\bar{1}} &= - c_{32} \bar{\xi}^3 \left( - 2 h_{12} + 3 c_{32} \left( 4 + r_\sg^3 R_1 \right) \right) / r_\sg^5 , \label{eq:k1AE1}
\end{align}
of $\tensor{\bar{\EuScript{E}}}{_t_t}$ from Eq.~\eqref{eq:fR2Ett}. Similarly, we obtain the coefficients
\begin{align}
	\oe_0 &= 0 \; , \; \oe_{12} = - 3 c_{32}^2 \bar{\xi}^{3/2} \left( 4 + r_\sg^3 R_1 \right) / r_\sg^5 , \label{eq:k1oe0oe12} \\
	\o_{32} &= - 2 c_{32} / r_\sg^5 , \label{eq:k1o32} \\
	\o_2 &= c_{32} \left( 2 h_{12} - 3 c_{32} \left( 4 + r_\sg^3 R_1 \right) / r_\sg^5 \right) . \label{eq:k1o2}
\end{align}
of $\tensor{\bar{\EuScript{E}}}{_t^r}$ and $\tensor{\bar{\EuScript{E}}}{^r^r}$ from Eqs.~\eqref{eq:fR2Etr}\textendash{}\eqref{eq:fR2Err}, where $R_1$ denotes the $x$ coefficient of the Ricci scalar $\bar{R} = \sum_j R_j x^j = 2/r_\sg^2 + R_1 x + \mathcal{O}(x^{3/2})$ and $R_{12} = 0$, cf.\ Eq.~\eqref{eq:k1Ricci}. With the expressions given in Eqs.~\eqref{eq:k1AE32}--\eqref{eq:k1o2}, it is easy to verify that both $k=1$ constraints Eq.~\eqref{eq:k1cond1} and Eq.~\eqref{eq:k1cond2} are satisfied identically.

\subsection{Compatibility with the $\boldsymbol{\lambda}$-expanded $\boldsymbol{k=0}$ class of black hole solutions} \label{subsec:Staro_lam_k0}

Equality of the coefficients in Eqs.~\eqref{con1b} and \eqref{con1t} follows in exactly the same fashion as in Sec.~\ref{subsec:Compf(R)k0}. Explicit calculation confirms that the coefficients of the MTG terms in the Starobinsky model Eqs.~\eqref{eq:fR2Ett}--\eqref{eq:fR2Err} obtained using the EMT expansion of Eq.~\eqref{eq:k0tau-non} and metric functions Eqs.~\eqref{eq:k0C-non}--\eqref{eq:k0h-non} coincide with those of Eq.~\eqref{eq:k0_con1_Staro} at the leading expansion order $\mathcal{O}(\lambda)$. Terms of order $\mathcal{O}(\lambda^0)$ vanish in accordance with Eq.~\eqref{eq:k0_con1_Staro} (note that $c_{12} \propto \mathcal{O}(\lambda)$). This confirms that the Starobinsky solution is consistent with the generic form of PBH solutions. However, since $\tilde{\Xi}_{(2)}$ is undetermined in the self-consistent approach, Eq.~\eqref{con1t} does not impose any constraints on the function $\xi$.

\section{Discussion}

We have analyzed the properties of metric MTG and derived several constraints that they must satisfy to be compatible with the existence of an apparent horizon. Since we have not specified the origin of the deviations from GR, the results presented here are generic and apply to all conceivable self-consistent metric MTG.

Constraints on a perturbative solution in a  particular metric MTG arise from two sources: first, the series expansions of the modified gravity terms $\tensor{\bar{\EuScript{E}}}{_\mu_\nu}$ in terms of the distance $x \defeq r - r_\sg$ from the horizon must follow a particular structure that is prescribed by the modified Einstein equations with terms that diverge in the limit $r \to r_\sg$. Second, a general spherically symmetric metric allows for two independent functions $C$ and $h$ that must satisfy three Einstein equations. The resulting relations between coefficients $\sigma_\ell$, $\omega_\ell$ of their perturbative corrections translate into relationships between the coefficients $c_\ell$ and $h_\ell$, and eventually components of the unperturbed EMT. These constraints must be satisfied identically. Otherwise, a valid solution of GR cannot be perturbatively extended to a solution of a MTG. Identities that must be satisfied for the existence of the perturbative $k=0$ solutions are given by Eqs.~\eqref{con0-1} and \eqref{con0-2}, and for the $k=1$ solution by Eqs.~\eqref{eq:k1cond1}--\eqref{eq:k1cond2}.

On the other hand, there are nonperturbative solutions that do not have a well-defined GR limit. In this case, the constraints on a MTG that are imposed by the existence of a regular apparent horizon formed in finite time of a distant observer are given by Eqs.~\eqref{con1b}--\eqref{con1t}.

Using the Starobinsky $R^2$ model, arguably the simplest possible MTG, we identify both perturbative and nonperturbative solutions. However, this is not the only theory that should be investigated: in a future article \cite{sM.2021}, we will consider generic $\mathfrak{f}(R)$ theories of the form $\mathfrak{f}(R) = R + \lambda \eF(R)$, where $\eF(R) = \varsigma R^q$ and $q, \varsigma \in \mathbb{R}$. In particular, this includes the case $q=1/2$ (i.e.\ $\mathfrak{f}(R) = R + \lambda \varsigma \sqrt{R}$) considered in Ref.~\cite{eE.gglN.sN.sdO.2020}, as well as the case of negative exponents $q<0$ considered in Ref.~\cite{smC.vD.mT.msT.2004}. More general MTG (e.g.\ those involving higher-order curvature invariants) will also be considered.

\acknowledgments
We thank Eleni Kontou, Robert Mann, Shin'ichi Nojiri, Vasilis Oikonomou, and Christian Steinwachs for useful discussions and helpful comments. SM is supported by an International Macquarie University Research Excellence Scholarship and a Sydney Quantum Academy Scholarship. The work of DRT was supported in part by the Southern University of Science and Technology, Shenzhen, China, and by the ARC Discovery project grant DP210101279.

\appendix

\section{Coefficients of perturbative corrections} \label{app:pertCoCoeff}

\subsection{$\boldsymbol{k=0}$ black hole solutions} \label{app:subsec:k0pertCoCoeff}

With the metric functions Eqs.~\eqref{eq:k0_metric_C}--\eqref{eq:k0_metric_h} of the $k=0$ solutions we obtain the following coefficients for the perturbative correction $\Sigma$ of Eq.~\eqref{eq:k0_correction_Sigma} from Eqs.~\eqref{eq:mGravEFEtt}--\eqref{eq:mGravEFErr}:
\begin{align}
	\sigma_{12 \vert tt} &= - \frac{r_\sg^{3/2} \ae_{\bar{1}}}{4 \sqrt{\pi} \bar{\xi} \bar{\Upsilon}} , \\
	\sigma_{12 \vert tr} &= - \frac{r_\sg^{3/2} \left( 4 \pi \tilde{\xi} \bar{\Upsilon}^2 + \sqrt{\bar{\xi}} \oe_{12} \right)}{2 \sqrt{\pi} \bar{\xi} \bar{\Upsilon}} , \\
	\sigma_{12 \vert rr} &= - \frac{r_\sg^{3/2} \o_0}{4 \sqrt{\pi} \bar{\Upsilon}} , \\
	\begin{split}
		\sigma_{1 \vert tt} &= \frac{4 \bar{\Upsilon} \ae_{\bar{1}} + 6 \sqrt{\pi} r_\sg^{3/2} \bar{\Upsilon}^2}{36 \pi \bar{\xi} \bar{\Upsilon}^3} \left( - \ae_{\overbar{12}} + 8 \pi \bar{\xi} (\tensor{\tilde{\tau}}{_t})_{12}
		\vphantom{8 \pi \bar{\xi} (\tensor{\tilde{\tau}}{_t})_{12} + \sqrt{\pi} r_\sg^{3/2} \ae_{\bar{1}} \left( 6 (\tensor{\bar{\tau}}{^r})_{12} - 5 (\tensor{\bar{\tau}}{_t})_{12} \right)}
		\right. \\
		& \hspace*{4mm} \left. + \sqrt{\pi} r_\sg^{3/2} \ae_{\bar{1}} \bigl( 6 (\tensor{\bar{\tau}}{^r})_{12} - 5 (\tensor{\bar{\tau}}{_t})_{12} \bigr) \right) ,
	\end{split} \label{eq:k1sigtt}
	\\
	\begin{split}
		\sigma_{1 \vert tr} &= - \frac{1}{12 \sqrt{\pi} \bar{\xi} \bar{\Upsilon}} \biggl[ \tilde{\xi} \left( - 4 \sqrt{\pi} \bar{\Upsilon} + 8 \pi r_\sg^{3/2} (\tensor{\bar{\tau}}{_t})_{12} \right) + 3 r_\sg^{3/2} \\
		& \times \sqrt{\bar{\xi}} \Bigl( - \oe_0 + 4 \pi \sqrt{\bar{\xi}} \bigl( (\tensor{\tilde{\tau}}{_t})_{12} + (\tensor{\tilde{\tau}}{^r})_{12} - 2 \bar{\Upsilon}^2 \omega_{12} \bigr) \Bigr)
		\biggr] ,
	\end{split} \label{eq:k1sigtr}
	\\
	\begin{split}
		\sigma_{1 \vert rr} &= \frac{1}{12 \pi \bar{\Upsilon}^3} \Bigl( - 2 \bar{\Upsilon} \o_0 + 6 \sqrt{\pi} r_\sg^{3/2} \bar{\Upsilon}^2 \bigl( - \o_{12} + 8 \pi (\tensor{\tilde{\tau}}{^r})_{12} \bigr) \\
		& + \sqrt{\pi} r_\sg^{3/2} \o_0 \bigl( (\tensor{\bar{\tau}}{_t})_{12} - 6 (\tensor{\bar{\tau}}{^r})_{12} \bigr) - 96 \pi^{3/2} r_\sg^{3/2} \bar{\Upsilon}^4 \omega_{12}
		\Bigr) .
	\end{split} \label{eq:k1sig1rr}
\end{align}

Via comparison of Eqs.~\eqref{eq:k1sigtr} and \eqref{eq:k1sig1rr} we can identify the coefficient
\begin{align}
	\begin{aligned}
		\omega_{12} &= \frac{1}{72 \pi \bar{\xi} \bar{\Upsilon}^2 r_\sg^{3/2}} \Bigl( 20 \sqrt{\pi} \tilde{\xi} \bar{\Upsilon} + 3 r_\sg^{3/2} \sqrt{\bar{\xi}} \oe_0 - 6 r_\sg^{3/2} \bar{\xi} \bigl( \o_{12} \\
		& \hspace*{-2mm} + 2 \pi (\tensor{\tilde{\tau}}{_t})_{12} - 6 \pi (\tensor{\tilde{\tau}}{^r})_{12} \bigr) + 16 r_\sg^{3/2} \pi \tilde{\xi} \bigl( 3 (\tensor{\bar{\tau}}{^r})_{12} - (\tensor{\bar{\tau}}{_t})_{12} \bigr)
		\Bigr)
	\end{aligned}
	\label{eq:k1w12}
\end{align}
for the perturbative correction $\Omega$ of Eq.~\eqref{eq:k0_correction_Omega}.

Substitution of Eq.~\eqref{eq:k1w12} into Eqs.~\eqref{eq:k1sigtr} and \eqref{eq:k1sig1rr} then yields
\begin{align}
	\begin{aligned}
		\sigma_{1 \vert tr} &= \sigma_{1 \vert rr} = \frac{1}{18 \sqrt{\pi} \bar{\xi} \bar{\Upsilon}} \\
		\left[ \vphantom{\sqrt{\bar{\xi}}} \right.
		& 3 r_\sg^{3/2} \sqrt{\bar{\xi}} \left( - 2 \oe_0 + \sqrt{\bar{\xi}} \bigl( \o_{12} + 8 \pi (\tensor{\tilde{\tau}}{_t})_{12} \bigr) \right) \\
		& - 4 \tilde{\xi} \left( 4 \sqrt{\pi} \bar{\Upsilon} + \pi r_\sg^{3/2} \bigl( 6 (\tensor{\bar{\tau}}{^r})_{12} - 5 (\tensor{\bar{\tau}}{_t})_{12} \bigr) \right)
		\left. \vphantom{\sqrt{\bar{\xi}}} \right] .
	\end{aligned}
	\label{eq:k1sig1trsig1rr}
\end{align}

Subsequent comparison of Eq.~\eqref{eq:k1sig1trsig1rr} and \eqref{eq:k1sigtt} establishes the relation Eq.~\eqref{con0-2} between the coefficients $\ae_{\overbar{12}}$, $\oe_0$, and $\o_{12}$.

\subsection{$\boldsymbol{k=1}$ black hole solution} \label{app:subsec:k1pertCoCoeff}

With the metric functions Eqs.~\eqref{eq:k1_metric_C}--\eqref{eq:k1_metric_h} of the $k=1$ solution we obtain the following coefficients for the perturbative correction $\Sigma$ of Eq.~\eqref{eq:k1_correction_Sigma} from Eqs.~\eqref{eq:mGravEFEtt} and \eqref{eq:mGravEFErr}:
\begin{align}
	\sigma_{32 \vert tt} &= r_\sg^2 \left( \frac{\ae_{\overbar{32}} r_\sg}{\bar{\xi}^3} - 8 \pi c_{32} \tilde{E} \right) , \\
	\sigma_{32 \vert rr} &= - \o_{32} r_\sg^3 + 8 \pi c_{32} r_\sg^2 \tilde{P} , \\
	\begin{split}
	\sigma_{2 \vert tt}
	&
	=
	\frac{r_\sg^2}{\bar{\xi}^3} \Bigl[
	\vphantom{8 \pi \bar{\xi}^3 \left( \left( c_2 - 3 c_{32}^2 \right) \tilde{E} - r_\sg \tilde{e}_2 \right)}
	\ae_{\overbar{32}} r_\sg \left( 3 c_{32} - 2 h_{12} \right) + \ae_{\bar{1}} r_\sg
	\\
	&
	\hspace*{9mm} - 8 \pi \bar{\xi}^3 \left( c_{32} \left( 3 c_{32} - h_{12} \right) \tilde{E} + r_\sg \tilde{e}_2 \right) \Bigr] ,
	\end{split} \label{eq:k1_sigma2_tt} \\
	\begin{split}
		\sigma_{2 \vert rr}
	&
		= - r_\sg^2 \Bigl( \o_2 r_\sg - c_{32} \big( 3 \o_{32} r_\sg
	\\ &
	\hspace*{9mm} - 8 \pi \left( 3 c_{32} + h_{12} \right) \tilde{P} \big) - 8 \pi r_\sg \tilde{p}_2 \Bigr) .
	\end{split} \label{eq:k1_sigma2_rr}
\end{align}

From the $x^0$ and $\sqrt{x}$ coefficients of Eq.~\eqref{eq:mGravEFEtr} we obtain
\begin{align}
	\tilde{E}_{\vert tr} &= \frac{\oe_0 r_\sg}{8 \pi c_{32} \bar{\xi}^{3/2}} , \label{eq:k1_tildeE_app} \\
	\tilde{\xi}_{\vert tr}
	&
	=- \frac{2 \bar{\xi} \sigma_{32}}{3 c_{32}} - \frac{4 r_\sg^3}{9 c_{32}^2 \sqrt{\bar{\xi}}} \left( \oe_{12} - 8 \pi \bar{\xi}^{3/2} \tilde{p}_2 \right)
	\label{eq:k1_chitil_tr} ,
\end{align}
where $\tilde{\Phi}=(\tilde{E}+\tilde{P})/2$ and $\tilde{\phi}_2 = (\tilde{e}_2 + \tilde{p}_2)/2$, see Eqs.~\eqref{eq:k1TildeTauRel32} and \eqref{eq:k1TildeTauRel}. By substitution of Eq.~\eqref{eq:k1_tildeE_app} into Eq.~\eqref{eq:k1_sigma2_tt} we obtain
\begin{align}
	\begin{aligned}
		\tilde{e}_2 + \tilde{p}_2 &= \frac{6 c_{32}^2 \tilde{\Phi}}{r_\sg} + \frac{1}{8 \pi \bar{\xi}^3} \Bigl[ \ae_{\bar{1}} + \ae_{\overbar{32}} \left( 3 c_{32} - 2 h_{12} \right)  \\
	& \hspace*{-4mm} + \bar{\xi}^3 \left( \o_2 - 3 c_{32} \o_{32} \right) + 2 \bar{\xi}^{3/2} \oe_0 \left( h_{12} - 3 c_{32} \right) \Bigr] .
	\end{aligned}
	\label{eq:sumTerm1}
\end{align}
from the comparison $\sigma_{2 \vert tt} - \sigma_{2 \vert rr} = 0$. Similarly, substitution of Eq.~\eqref{eq:k1_chitil_tr} into Eq.~\eqref{eq:k1sigma32R} and subsequent comparison of $\sigma_{32 \vert R} - \sigma_{32 \vert tt} = 0$ yields
\begin{align}
	\tilde{e}_2 + \tilde{p}_2 &= \frac{6 c_{32}^2 \tilde{\Phi}}{r_\sg} + \frac{3 \ae_{\overbar{32}} c_{32} - \bar{\xi}^{3/2} \left( 6 c_{32} - \oe_0 \oe_{12} \right)}{4 \pi \bar{\xi}^3} .
	\label{eq:sumTerm2}
\end{align}
Subtracting Eq.~\eqref{eq:sumTerm1} from Eq.~\eqref{eq:sumTerm2} and subsequent multiplication by $8 \pi \bar{\xi}^3 r_\sg$ yields
\begin{align}
	\begin{aligned}
		- r_\sg \Bigl[ \ae_{\bar{1}} & - \ae_{\overbar{32}} \left( 3 c_{32} + 2 h_{12} \right) + \bar{\xi}^{3/2} \left( \bar{\xi}^{3/2} \left( \o_2 - 3 c_{32} \o_{32} \right) \right. \\
		& + 6 c_{32} \oe_0 + 2 h_{12} \oe_0 - 2 \oe_{12}
		\left. \vphantom{\bar{\xi}^{3/2}}
		\right) \Bigr] = 0 .
	\end{aligned}
	\label{eq:subExpr}
\end{align}
Lastly, substituting $\ae_{\overbar{32}}$ from Eq.~\eqref{eq:k1cond1} into \eqref{eq:subExpr} and rearranging gives Eq.~\eqref{eq:k1cond2}.

\onecolumngrid
\section{Additional explicit expressions for $\boldsymbol{k=0}$ black hole solutions} \label{app:k0expressions}

\vspace{5mm}
Explicit expressions for the individual terms in Eq.~\eqref{eq:k0con2Ricci}:
\vspace{3mm}
{\small
\begin{align}
	\begin{split}
		(\tensor{\bar{R}}{_t_t})_{\overbar{12}} &= 2 \sqrt{\bar{\xi}} (\tensor{\bar{R}}{_t^r})_0 - \bar{\xi} (\tensor{\bar{R}}{^r^r})_{12} \\
		&= - \frac{1}{24 c_{12} r_\sg^3} \Biggl[ - 2 c_{12}^3 \bar{\xi} \Bigl( h_{12}^3 r_\sg + 6 h_{32} r_\sg + h_{12} \left( 9 h_1 r_\sg - 6 \right) \Bigr) 6 r_\sg \sqrt{\bar{\xi}} \left( c_1 - 1 \right) \left( \sqrt{\bar{\xi}} \left( c_1 - 1 \right)^2 - 2 r_\sg c_{12}^\prime \vphantom{\sqrt{\bar{\xi}}} \right) \\
		& \hspace{23mm} \times \biggl[ 4 c_{32} r_\sg \bar{\xi} \left( c_1 - 1 \right)  + r_\sg \sqrt{\bar{\xi}} \left( 2 r_\sg c_1^\prime - h_{12} \left( \sqrt{\bar{\xi}} \left( c_1 - 1 \right)^2 - 2 r_\sg c_{12}^\prime  \vphantom{\sqrt{\bar{\xi}}} \right) \right) \biggr] \\
		& \hspace{23mm} + 6 c_{12}^2 \biggl( \bar{\xi} \Bigl( 3 - 2 h_1 r_\sg + c_{32} h_{12} r_\sg - h_{12}^2 r_\sg + c_1 \left( -3 + 2 h_1 r_\sg  + h_{12}^2 r_\sg \right) \Bigr) + r_\sg^2 \sqrt{\bar{\xi}} h_{12}^\prime \biggr) \Biggr] .
		\label{eq:k0Rttb12}
	\end{split}
	\\
	\begin{split}
		(\tensor{\bar{R}}{_t^r})_0 &= c_{12} \sqrt{\bar{\xi}} \left( c_1 - 1 \right) /  r_\sg^3 .
		\label{eq:k0Rtr0}
	\end{split}
	\\
	\begin{split}
		(\tensor{\bar{R}}{^r^r})_{12} &= \frac{1}{24 c_{12} r_\sg^3 \sqrt{\bar{\xi}}} \Biggl[ - 2 c_{12}^3 \sqrt{\bar{\xi}} \Bigl( h_{12}^3 r_\sg + 6 h_{32} r_\sg + h_{12} \left( 9 h_1 r_\sg - 6 \right) \Bigr) + 6 r_\sg \left( c_1 - 1 \right) \left( \sqrt{\bar{\xi}} \left( c_1 - 1 \right)^2 - 2 r_\sg c_{12}^\prime \right) \\
		 & \hspace{23mm} + 3 c_{12} r_\sg \biggl[ 4 c_{32} \sqrt{\bar{\xi}} \left( c_1 - 1 \right) + 2 r_\sg c_1^\prime - h_{12} \left( \sqrt{\bar{\xi}} \left( c_1 - 1 \right) - 2 r_\sg c_{12}^\prime \right) \biggr] \\
		  & \hspace{23mm} + 6 c_{12}^2 \biggl[ \sqrt{\bar{\xi}} \left( h_{12} c_{32} r_\sg - 2 h_1 r_\sg -h_{12}^2 r_\sg + c_1 \left( 5 + 2 h_1 r_\sg + h_{12}^2 r_\sg \right) - 5 \right) + r_\sg^2 h_{12}^\prime \biggr] \Biggr] .
		  \label{eq:k0Rrr12}
	\end{split}
\end{align}
}
\vspace{5mm}

Explicit expressions for the MTG coefficients $\ae_{\overbar{12}}$, $\oe_0$, $\o_{12}$ in the Starobinsky model of the $k=0$ solution (see Subsec.~\ref{subsec:Compf(R)k0}):
\vspace{2.5mm}
{\small
\begin{align}
	\begin{split}
	\ae_{\overbar{12}} & = \frac{1}{12 c_{12} r_\sg^5} \Biggl[ - 12 r_\sg \bar{\xi} R_0 \left( c_1 - 1 \right)^3 + 2 c_{12}^3 \bar{\xi} \Bigl( 2 R_0 \bigl( - 6 h_{12} + r_\sg ( 9 h_{12} h_1 + h_{12}^3 + 6 h_{32} ) \bigr) - 6 h_{12} r_\sg^3 R_1 + 3 r_\sg^2 R_{12} \\
	& \hspace{23mm} \times \Bigl( - 6 + r_\sg \left( h_{12}^2 + h_1\right) \Bigr) + 24 r_\sg^2 R_0 \sqrt{\bar{\xi}} c_{12}^\prime \left( c_1 - 1 \right) - 3 c_{12} r_\sg \sqrt{\bar{\xi}} \biggl[ 8 c_{32} \sqrt{\bar{\xi}} R_0 \left( c_1 - 1 \right) + 2 h_{12} R_0 \\
	& \hspace{23mm} \times \Bigl( 2 r_\sg c_{12}^\prime \sqrt{\bar{\xi}} \left( c_1 - 1 \right)^2 \Bigr) + r_\sg \Bigl( 4 R_0 c_1^\prime + 2 r_\sg^2 R_{12} c_{12}^\prime - 4 R_0^\prime \left( c_1 -  1 \right) - r_\sg \sqrt{\bar{\xi}} R_{12} \left( c_1 - 1 \right)^2 \Bigr) \biggr] \\
	& \hspace{23mm} + 6 c_{12}^2 \biggl[ -2 \bar{\xi} R_0 \biggl( 5 + r_\sg \Bigl( 2 h_1 \left( c_1 - 1 \right) + h_{12} \Bigl( c_{32} + h_{12} \left(  c_1 - 1 \right) \bigr) \Bigr) - 5 c_1 \biggr) + 4 \bar{\xi} R_0^2 \\
	& \hspace{23mm} + r_\sg^3 \bar{\xi} \Bigl( R_{12} \left( h_{12} - h_{12} c_1 + c_{32} \right) + 4  R_1 \left( c_1 - 1 \right) \Bigr) - 2 r_\sg^2 \sqrt{\bar{\xi}} R_0 h_{12}^\prime - r_\sg^4 \sqrt{\bar{\xi}} R_{12}^\prime \biggr] \Biggr] .
	\label{eq:k0aeb12}
	\end{split}
	\\
	\begin{split}
		\oe_0 &= \frac{1}{2 r_\sg^5} \Biggl[ c_{12}^2 \sqrt{\bar{\xi}} \Bigl( 2 h_{12} \left( 4 R_0 + r_\sg^3 R_1 \right) + r_\sg^2 \left( 2 h_1 r_\sg - 3 \right) \Bigr) - r_\sg^4 R_{12} c_{12}^\prime \\
		& \hspace{23mm} + c_{12} \biggl( 8 \sqrt{\bar{\xi}} R_0 \left( c_1 - 1 \right) + r_\sg^2 \Bigl(  r_\sg \sqrt{\bar{\xi}} \left( c_1 - 1 \right) \left( 2 R_1 - h_{12} R_{12} \right)  + 4 h_{12} R_0^\prime - 3 r_\sg^2 R_{12}^2 \Bigr) \biggr) \Biggr] .
		\label{eq:k0oe0}
	\end{split}
	\\
	\begin{split}
		\o_{12} &= \frac{1}{12 c_{12} r_\sg^5 \sqrt{\bar{\xi}}} \Biggl[ - 2 c_{12}^3 \sqrt{\bar{\xi}}\biggl( 2 R_0 \Bigl( r_\sg \left( 9 h_{12} h_1 + h_{12}^3 + 6 h_{32} \right) - 30 h_{12} \Bigr) -18 h_{12} r_\sg^3 R_1 + 3 \left( - 3 h_1 + h_{12}^2 \right) \biggr) \\
		 & \hspace{23mm} + 12 r_\sg R_1 \left( c_1 - 1 \right) \left( \sqrt{\bar{\xi}} \left( c_1 - 1 \right)^2 - 2 r_\sg c_{12}^\prime \right) - 3 c_{12} r_\sg \biggl[ - 8 c_{32} \sqrt{\bar{\xi}} R_0  \left( c_1 - 1 \right) + 2 h_{12} R_0  \\
		 & \hspace{23mm} \times \left( \sqrt{\bar{\xi}} \left( c_1 - 1 \right)^2 - 2 r_\sg c_{12}^\prime \right) r_\sg \biggl( r_\sg \sqrt{\bar{\xi}} R_{12} \left( c_1 - 1 \right)^2 - 4 R_0 c_1^\prime + 2 r_\sg^2 R_{12} c_{12}^\prime + 4 R_0^\prime \left( c_1 - 1 \right) \biggr) \biggr] \\
		 & \hspace{23mm} - 6 c_{12}^2 \biggl[ - 2 \sqrt{\bar{\xi}} R_0 \Bigl( 3 \left( c_1 - 1 \right) \Bigr) + r_\sg \Bigl( 2 h_1 \left( c_1 - 1 \right) + h_{12} \bigl( c_{32} + h_{12} \left( c_1 - 1 \right) \bigr) \Bigr) \\
		 & \hspace{23mm} + 4 \sqrt{\bar{\xi}} R_0^2 - 2 r_\sg^2 R_0 h_{12}^\prime + r_\sg^2 \biggl( c_{32} r_\sg \sqrt{\bar{\xi}} R_{12} + h_{12} \Bigl( r_\sg \sqrt{\bar{\xi}} R_{12} \left( c_1 - 1 \right) - 8 R_0^\prime \Bigr) + 5 r_\sg^2 R_{12}^\prime \biggr) \biggr] \Biggr] .
		 \label{eq:k0o12}
	\end{split}
\end{align}
}

\twocolumngrid

\end{document}